\documentclass[parskip=full,12pt]{article}

\usepackage{amssymb}
\usepackage{amsthm}
\usepackage{amsmath}
\usepackage{float,booktabs}
\usepackage{graphicx}
\usepackage{natbib}
\usepackage{url}
\usepackage{color,soul}
\usepackage{array,multirow}
\usepackage[top=1in, bottom=1.3in, left=0.875in, right=0.875in]{geometry}
\usepackage[linesnumbered,ruled]{algorithm2e}
\usepackage[labelformat=simple]{subcaption}
\usepackage{bm}
\usepackage[linesnumbered,ruled]{algorithm2e}
\usepackage{caption}

\def\spacingset#1{\renewcommand{\baselinestretch}
{#1}\small\normalsize} \spacingset{1}

\newcolumntype{P}[1]{>{\centering\arraybackslash}p{#1}}
\newcommand*{\myfont}{\fontfamily{lmss}\selectfont}
\DeclareTextFontCommand{\textpython}{\myfont}

\DeclareCaptionLabelFormat{boldnumber}{\bothIfFirst{#1}{~}\textbf{#2}}
\title{Cost-sensitive Multi-class AdaBoost for Understanding Driving Behavior with Telematics}

\author{Banghee So \thanks{Department of Mathematics, University of Connecticut, 341 Mansfield Road, Storrs, CT, 06269-1009, USA. Email: \texttt{banghee.so@uconn.edu}.} 
\and Jean-Philippe Boucher\thanks{D\'{e}partement de Math\'{e}matiques, Universit\'{e} du Qu\'{e}bec \`{a} Montr\'{e}al, 201, Avenue du Pr\'{e}sident-Kennedy, Montr\'{e}al, Qu\'{e}bec, H2X 3Y7, Canada. Email: \texttt{boucher.jean-philippe@uqam.ca}.}
\and Emiliano A. Valdez\thanks{Corresponding author; Department of Mathematics, University of Connecticut, 341 Mansfield Road, Storrs, CT, 06269-1009, USA. Email: \texttt{emiliano.valdez@uconn.edu}.}}

\begin{document}

\maketitle

\begin{abstract}

Powered with telematics technology, insurers can now capture a wide range of data, such as distance traveled, how drivers brake, accelerate or make turns, and travel frequency each day of the week, to better decode driver's behavior. Such additional information helps insurers improve risk assessments for usage-based insurance (UBI), an increasingly popular industry innovation. In this article, we explore how to integrate telematics information to better predict claims frequency. For motor insurance during a policy year, we typically observe a large proportion of drivers with zero claims, a less proportion with exactly one claim, and far lesser with two or more claims. We introduce the use of a cost-sensitive multi-class adaptive boosting (AdaBoost) algorithm, which we call \texttt{SAMME.C2}, to handle such imbalances. To calibrate \texttt{SAMME.C2} algorithm, we use empirical data collected from a telematics program in Canada and we find improved assessment of driving behavior with telematics relative to traditional risk variables. We demonstrate our algorithm can outperform other models that can handle class imbalances: \texttt{SAMME}, \texttt{SAMME} with \texttt{SMOTE}, \texttt{RUSBoost}, and \texttt{SMOTEBoost}. The sampled data on telematics were observations during 2013-2016 for which 50,301 are used for training and another 21,574 for testing. Broadly speaking, the additional information derived from vehicle telematics helps refine risk classification of drivers of UBI.

\vspace{0.8cm}

\noindent \textbf{Keywords}: Vehicle telematics, usage-based insurance, cost-sensitive learning, Adaboost, \texttt{SMOTE}, \texttt{SAMME}, \texttt{SAMME.C2} 

\end{abstract}

\newpage

\section{Introduction} \label{sec:intro}

Telematics refers to the use of telecommunication devices and technology to transmit and store information. It is derived from the French word, t\'{e}l\'{e}matique,  which combines the words ``telecommunications'' and ``computing science.'' We are seeing a growing list of applications of telematics technology in many diverse fields: a radio receiver installed in a drone for journalism reporting or for private investigation, a smart home system that remotely controls temperature, lighting, appliance, and even alarms for security, an electronic system for better communication between healthcare professionals and their patients in health telematics (\citet{wong1998}), and most predominantly, a device installed in a car or a mobile app to remotely monitor driving habits for insurance rating.

While the use of telematics-based insurance has not fully matured in the insurance industry, it is gaining more attraction to drivers, especially to younger ones, which gives them opportunity to save on premiums in exchange for driving behavior being monitored. According to \citet{cipr2015}, Progressive Insurance Company offered the first such UBI in the early 2000s where the company, in collaboration with General Motors, began to offer premium discounts linked to monitoring of driving activities and behavior. A tracking device, upon driver's agreement, is installed in the car to collect the information through GPS technology. What follows next is the acceleration in technology advancement leading to some variations in UBI with similar names floating around that include, for example, Pay as you Drive (PAYD), Pay how you Drive (PHYD), Pay as you Drive as you Save (PAYDAYS), Pay per mile, and Pay as you Go (PASG). The use of telematics for UBI enables insurers to collect metrics of driving habits to help improve driver risk profile, thereby allowing for a more flexible pricing mechanism to reflect this observed behavior. 

There is a positive social effect to vehicle telematics: it encourages better driving behavior. UBI programs have many potential benefits to consumers, insurers, and the society, in general. It allows insurers to put a price tag that links more closely to individual's driving habits. This helps insurers to increase the predictability of their profit margin and gives more careful drivers the opportunity for more affordable premiums. Consumers can control premium costs by adopting safer driving habits or reducing frequency of driving. Moreover, it benefits the society because with safer driving and fewer drivers on the road, this reduces accidents, congestion, and car emissions. In order to get the optimal benefits of UBI to both insurers and their policyholders, it becomes subsequently crucial to identify the more significant telematics variables that truly affects the occurrence of car accidents.

It is comforting to know that there is a growing literature on telematics in actuarial science and insurance; many of these research works find additional value from telematics for better claims predictions, risk classification, and premium assessments. The work of \citet{ayuso2016gender} uses, interestingly, survival models to conclude that gender discrimination is unnecessary in the presence of sufficient telematics information on driving behavior.  \citet{boucher2017gam} offers the benefits of using generalized additive models (GAM) to gain additional insights as to how premiums can be more dynamically assessed for PAYD policies based on telematics information. Using zero-inflated Poisson regression models, \citet{guillen2019telem} investigated how telematics information helps explain part of the occurrence of zero accidents not typically accounted by traditional risk factors. For tariff determination, \citet{ayuso2019auto} uses classical frequency models to incorporate significant information drawn from telematics metrics using data from a portfolio of PAYD policies issued by a Spanish insurance company. Additional research that have revealed the benefits of telematics metrics for improved understanding of driving behavior includes, but not limited to, \citet{constantinescu2018impact}, \citet{verbelen2018telem}, \citet{perez2019quantile}, \citet{pesantez2019xgboost}, and \citet{guillen2020telem}.

We distinguish our work from previous literature by addressing the sparsity of recorded claims for datasets with telematics information, leading to a highly imbalanced observed accident frequencies. For most portfolios of motor insurance, it is characteristically often a rare event to observe exactly one claim from a policy, let alone observing two or more claims. As we shall see, this is even more noticeable in the case of telematics data because there is a perceived self-selection for drivers with UBI. Because of the attractiveness of potential premium savings, policyholders of UBI believe and tend to be more careful drivers making the sparsity an even more important issue. The early work of \citet{pednault2000ins} somehow addresses this highly imbalanced classification suggesting the use of a tree-based model with the log-likelihood as an impurity function for identifying the splitting of the regions; the common impurity measure used for classification trees is the Gini index. The work lacks the empirical evidence to support the method.

The proposed algorithm in this paper is motivated by our telematics dataset drawn from a UBI program offered by a Canadian-owned insurance cooperative. We focus on the accident frequency, that is, the number of accidents or claims made during our observation period, as the response variable. According to our training data, we have 97.1\% observations with zero claims, 2.8\% with exactly one claim, and merely 0.1\% with two or more claims. Our observations are clearly highly imbalanced, and yet we want to investigate the predictive power of using telematics metrics, apart from the traditional metrics, to understand and predict claims frequency.

We employ an innovative multi-class classification algorithm, for which we refer to as \texttt{SAMME.C2} that can handle such an imbalance. The proposed algorithm has its origin from the work of \citet{zhu2009mclass}, which introduced \texttt{SAMME} (\texttt{S}tagewise \texttt{A}dditive \texttt{M}odeling using \texttt{M}ulti-class \texttt{E}xponential loss function), a multi-class AdaBoost classification model. Adaptive boosting (AdaBoost), as introduced by \citet{freund1997decision}, is an iterative classification algorithm that combines several weak and inaccurate learners to improve prediction accuracy. In contrast with other similar AdaBoost techniques, \texttt{SAMME.C2} uses cost-sensitive learning mechanism to rebalance and tilt the class distribution by accounting for the costs of prediction errors. This paper evaluates the competitiveness of our proposed algorithm to other models that have appeared in the machine learning literature and that have demonstrated to handle imbalanced classes.

In particular, using both simulated and real datasets, we find that \texttt{SAMME.C2} algorithm outperforms other multi-class classification models that can handle class imbalances: \texttt{SAMME}, \texttt{SAMME} with \texttt{SMOTE}, \texttt{RUSBoost}, and \texttt{SMOTEBoost}. Each of these methods is briefly explained in the subsequent section. While performance statistics such as Recall and Precision are not reasonable measures for comparison, we chose to use the G-mean metric for comparison. This performance metric is simply a geometric average of recall statistics for all classes. In addition, we find that relatively telematics variables are more significantly important predictors than traditional variables of accident frequencies.

For the rest of this paper, it has been organized as follows. Section 2 provides an overview of related works on adaptive boosting learning algorithms suitable to handle class imbalances. In Section 3, we introduce our new \texttt{SAMME.C2} algorithm, which is largely based on some integration of \texttt{SAMME} and \texttt{Ada.C2} methods; these two latter methods are therefore described in detail. We also briefly explain various performance metrics used to compare various classification models. Section 4 presents the simulation results to assess the performance of \texttt{SAMME.C2}. Section 5 presents results based on our telematics dataset. Section 6 concludes the paper.   

\section{Related work} \label{sec:related}

In this section, we give an overview of related work and techniques, which may be categorized into three levels: (a) Data level; (b) Algorithm level; and (c) Cost-sensitive learning level.

\subsection{Handling minority classes} \label{sub:imb}

As pointed out by \citet{yang200610probs}, one of the challenges of data mining is dealing with observations that may suffer from the presence of imbalanced classes. At the data level, as with our telematics dataset, the class distribution is inherently not balanced creating classes that are considered majority (too many) or minority (too few). It becomes difficult to predict classes belonging to the minority since there are only a few samples to learn from regarding the features to predict from this class. One obvious approach is to resample from your dataset so that the class distribution is rebalanced by over-sampling (or under-sampling) from the underrepresented (or overrepresented) classes. The increasingly popular over-sampling approach is the technique developed by \citet{chawla2002smote} called \texttt{SMOTE} (Synthetic Minority Over-sampling Technique). In contrast to over-sampling with replication, the \texttt{SMOTE} algorithm creates synthetic samples for the minority class. This is generated by drawing samples using the method of $k$ nearest neighbors (KNN) and linearly connecting them to produce the synthetic samples. As concluded by \citet{chawla2002smote}, it ``works to cause the classifier to build larger decision regions that contain nearby minority class points.'' While \texttt{SMOTE} is therefore utilized to increase prediction accuracy over minority classes, we will examine this algorithm in combination with boosting methods that are utilized to increase accuracy over the entire data.

\subsection{Boosting} \label{sub:imb}

At the algorithm level, boosting techniques are believed to be one of the most powerful learning algorithms discovered in recent years. Boosting is based on the principle of combining several weak learners to produce a strong learner for more improved and accurate predictions. Starting with equal observation weights, the algorithm is an iterative process of fitting classifiers at each step and adjusting the weights, in the subsequent steps, according to the result of the classification. More weights are given to observations that have been misclassified. Note that, as pointed out by \citet{hastie2009}, while boosting was originally intended for classification problems, the idea has been expanded to regression problems.

The first practical boosting algorithm was introduced by \citet{freund1997decision} and is referred to as \texttt{AdaBoost.M1}. AdaBoost is now the name that refers to a class of adaptive boosting algorithms. Assume that we are given a set of data denoted by $(\bm{x}_i,y_i)$ for $i=1,\ldots,m$ where $\bm{x}_i$ is a set of feature variables and $y_i$ is a binary variable. \texttt{AdaBoost.M1} is a well-known iterative boosting algorithm. Beginning with a distribution of equal weights to the observations, this is updated after each iteration based on $\alpha_t$, which is a function of the weighted classification error
\begin{equation} \label{eq:1}
\epsilon_t = \frac{\sum_{i=1}^{m} D_i I(y_i \ne h_t(\bm{x}_i))}{\sum_{i=1}^m D_i},
\end{equation}
where $D_i$ is the distribution of the weights and $h_t(\bm{x}_i)$ is the classifier at step $t \in {1,2,\ldots,T}$. This weighted error tends to increase during the iteration process while the final classifier's training error gradually decreases. It has been shown (\citet{friedman2000addlog} and \citet{hastie2009}) that \texttt{AdaBoost.M1} is equivalent to an additive model with a minimization of an exponential loss function and therefore belongs to the traditional statistical family of Forward Stagewise Additive Models. This viewpoint helps the algorithm to be efficient and to have straightforward statistical interpretation. Several variants of adaptive boosting algorithms have appeared in the literature. See \citet{ferreira2012review} for a review.

AdaBoost algorithms have gained widespread popularity and several works reveal advantages of the algorithms. It has been shown that in \citet{schapire1999improv}, the training error of the final classifier is bounded, and that in \citet{freund1997decision}, if each weak classifier is slightly better than random, then the training error drops exponentially fast in $T$,  the number of weak classifiers. Some have also reported that AdaBoost is robust to overfitting showing that test error consistently decreases and then levels off as more classifiers are added. In terms of practicality, many empirical applications in machine learning have shown that AdaBoost algorithms are superior classifiers. For instance, \citet{friedman2000addlog} called AdaBoost with decision trees the ``best off-the-shelf classifier in the world.''

\texttt{AdaBoost.M1} has been extended to multi-class classification problems where $y_i$ belongs to the set $\{1,2,\ldots, K\}$. The extension \texttt{AdaBoost.M2} in \citet{freund1997decision}, which is based on a pseudo-loss function instead of the error rate, is suitable for handling multi-class problems. The \texttt{AdaBoost.MH} developed by \citet{schapire1999improv} is an adaptive boosting multi-class algorithm that is based on the Hamming loss function. Because this loss function is applied to create a set of binary problems, the procedure may be slow and thereby inefficient. These are just a few extensions, but in this paper, we will compare our proposed method to the multi-class extension of the forward stagewise additive model, which is based on the generalization of exponential loss to multi-class given by
\begin{equation}
L(\bm{y}_i,\bm{f}(\bm{x}_i)) = \exp\!\Big(\!-\frac{1}{K}\sum_{k=1}^K y_k f_k(\bm{x})\Big) = \exp\!\Big(\!-\frac{1}{K}\bm{y}'_i\bm{f}(\bm{x}_i)\Big),\end{equation}
for observation $i$. Developed by \citet{zhu2009mclass}, this is what has been referred to as the \texttt{SAMME} algorithm and detailed steps of the algorithm are summarized in Appendix A.

It is quite suitable to combine the benefits of resampling and boosting, and therefore, we also considered the following algorithms:
\begin{itemize}
\setlength\itemsep{0.09em}
\item \texttt{SAMME} with \texttt{SMOTE} sampling;
\item \texttt{SMOTEBoost}, described in \citet{chawla2003smboost}, is an approach for learning from minority classes based on a combination of \texttt{SMOTE} and \texttt{AdaBoost.M2}; and
\item \texttt{RUSBoost}, described in  \citet{seiffert2010rusboost}, is an algorithm that has the same goal as \texttt{SMOTEBoost} but replaces \texttt{SMOTE} sampling with random undersampling.
\end{itemize}

\subsection{Cost-sensitive learning} \label{sub:cost}

Cost-sensitive learning provides for an additional layer of complexity in the algorithm to further improve prediction accuracy. In particular, it takes into account misclassification costs by adding a penalty to predictions that lead to incorrect classification. While costs added are primarily those accounting for misclassification, it has been identified that cost adjustments may be made to reflect other types such as computational efficiency/complexity, data collection, or model evaluation. The primary objective is to minimize the total costs of the model. Cost-sensitive algorithms that minimizes misclassification costs in classification problems first appeared in \citet{pazzani1994reduce}. The cost matrix is an additional input to the learning procedure and is also used to evaluate the ability of the learned procedure to reduce misclassification costs. In the context of adaptive boosting, the cost adjustment function is used to modify the updating of the weights at each iteration. See \citet{galar2012review}. The most familiar method of AdaBoost combined with cost-sensitive learning is \texttt{Ada.C2}, which inspired our algorithm suggested in this paper. In the next section, we discuss this algorithm in detail that gives us a prelude to our proposed multi-class cost-sensitive algorithm.

\section{The \texttt{SAMME.C2} algorithm} \label{sec:sc2}

This algorithm combines the benefits of boosting and cost-sensitive algorithms for handling class imbalances in multi-class classification problems. Boosting algorithms are generally considered advantageous because implementation is straightforward, methods are statistically justified and generally suitable for many kinds of classification problems, and there are fewer issues with overfitting. By directly penalizing misclassified samples, cost-sensitive learning algorithms provide the added benefits of prediction accuracy especially for minority classes. The cost-sensitive component of our proposed algorithm is inspired by the \texttt{Ada.C2}.

\subsection{\texttt{Ada.C2} method} \label{sub:adac2}

AdaBoost models treat samples of different classes equally. This is because weights of misclassified samples from different classes are increased by an identical ratio and weights of correctly classified samples from different classes are decreased by another identical ratio.  A desirable boosting strategy for imbalance dataset is one that is able to distinguish different classes of samples and boost more weights on the samples associated with higher costs. The concept of \texttt{Ada.C2} was introduced by \citet{sun2007cost}. The method adds the cost item to each sample as follows: input data to the algorithm consists of $(\bm{x}_i, y_i, \text{Cost}_i)$, for $i=1,2,\ldots,N$ where $N$ is the total number of training samples. The differences from the original AdaBoost are
\begin{enumerate}
\setlength\itemsep{0.09em}
\item The updating of the distribution of the dataset at each iteration has the form:
\[
D_{t+1}(i) = \dfrac{\text{Cost}_i \, D_t(i) \exp(-\alpha_t I(y_i = h_t(\bm{x}_i)))}{\sum_{j=1}^{N} \text{Cost}_j \, D_t(j) \exp(-\alpha_t I(y_j = h_t(\bm{x}_j)))}
\]
\item The weight of each classifier is:
\[
\alpha_t = \frac{1}{2} \log\! \left(\dfrac{\sum_{i=1}^N \text{Cost}_i \, D_t(i) \, I(y_i =h_t(\bm{x}_i))}{\sum_{i=1}^N  \text{Cost}_i \, D_t(i) \, I(y_i \ne h_t(\bm{x}_i))} \right)
\]
\end{enumerate}

AdaBoost is accuracy-oriented and therefore, its weighting strategy may still tilt towards the majority class since it contributes more to the overall classification accuracy. However, \texttt{Ada.C2} strategy of updating data distribution is by giving higher costs to minority class. On one hand, when minority classes are misclassified,  the weights increase more than when majority classes are misclassified. On the other hand, when minority classes are correctly classified, the weights decrease more conservatively than when majority classes are correctly classified. Subsequently the algorithm achieves an ideal weighting strategy to address the issue of highly imbalanced classes. The steps in the algorithm are summarized in Algorithm \ref{alg:adac2}.

\subsection{The proposed algorithm} \label{sub.sammec2}

The proposed algorithm for which we call \texttt{SAMME.C2} is a blend of \texttt{SAMME} and \texttt{Ada.C2} algorithms. Although we inherit all the algorithmic steps in \texttt{Ada.C2}, there are importance differences. First, we inherit the calculation formula of $\alpha_t$ at the iterative step $t$ from the \texttt{SAMME} algorithm that includes the addition of $\log(K-1)$. As pointed out by \citet{zhu2009mclass}, this adjustment term is crucial for multi-class classification problems as it strengthens the assurance of ``the accuracy of each weak classifier to be better than random guessing.'' It can be shown that the presence of the term $\log(K-1)$ arises as a consequence of the solution to the optimization based on the extended muti-class exponential loss function. Second, the calculation of the weighted classification error, $\epsilon_t$, is not adjusted with the cost values in our algorithm. This is a result necessary to prove our algorithm follows the forward stagewise additive model.

The steps in the algorithm are summarized below:

\begin{algorithm}[H]
\KwIn{Training dataset $\ \bm{x}_i \in X$, $y_i \in Y = \{1,2,\ldots,K\}$, $\text{Cost}_i \in (0,1]$, $T$}
\KwOut{Final classifier $\ H(\bm{x}_i)$}
Set initial distribution of dataset equally distributed:  $D_1(i) = \frac{1}{N}, \quad i =1,2,\ldots,N$ \;
\For{$t=1, \ldots, T$}{
Train weak classifier using the distribution $D_t$\;
Get weak classifier $h_t: X \rightarrow k \in \{1,2,\ldots,K\}$ \;
Compute $\epsilon_t = \dfrac{\sum_{i=1}^{N} D_t(i) I(y_i \ne h_t(\bm{x}_i))}{\sum_{i=1}^N D_t(i)}$ \;
Choose $\alpha_t = \frac{1}{2} \log\!\Big(\dfrac{1-\epsilon_t}{\epsilon_t}\Big) + \log(K-1)$ \;
Update $D_{t+1}(i) = \dfrac{\text{Cost}_i \, D_t(i) \exp(-\alpha_t I(y_i = h_t(\bm{x}_i)))}{\sum_{j=1}^{N} \text{Cost}_j \, D_t(j) \exp(-\alpha_t I(y_j = h_t(\bm{x}_j)))}$ \;
}
Return the final classifier: $\ H(\bm{x}_i) = \underset{k}{\mathrm{argmax}} {\  \sum_{t=1}^{T} \alpha_t I(h_t(\bm{x}_i) = k)}$ \;
\caption{\texttt{SAMME.C2}: cost-sensitive multi-class AdaBoost}\label{alg:sammec2}
\end{algorithm}

\begin{figure}[htbp]
\centering
\includegraphics[scale=0.475]{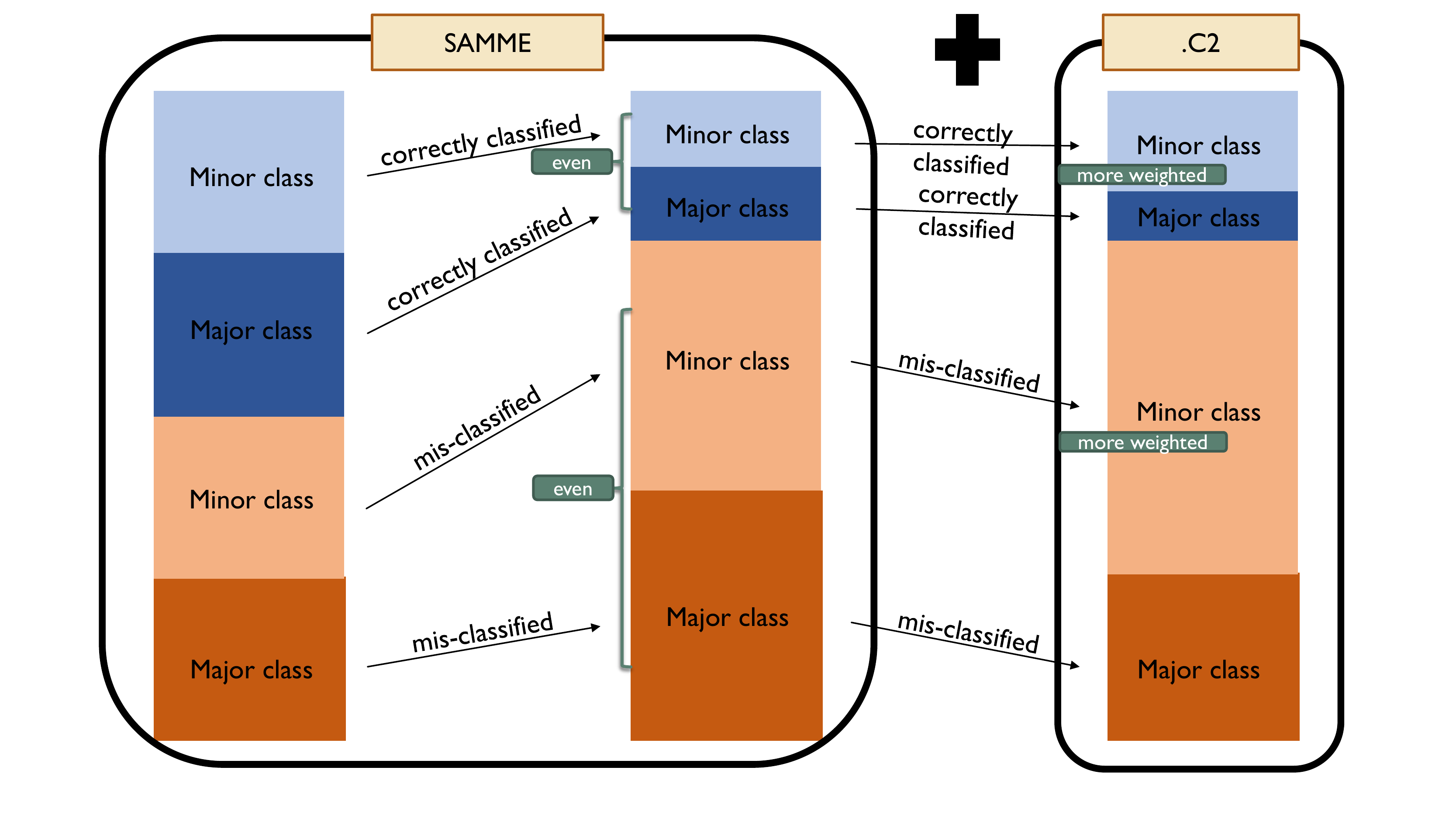}
\caption{Visualizing the effect of the \texttt{SAMME.C2} algorithm on classifying majority/minority class.} \label{fig1:sammec2}
\end{figure}

Figure \ref{fig1:sammec2} displays a visual representation of the comparative effect of how the algorithm correctly classifies (or misclassifies) majority and minority classes. In the \texttt{SAMME} algorithm, there is an even redistribution of correct classification (or misclassification) for both majority and minority classes. On the other hand, with the addition of a cost-sensitive learning mechanism, this redistribution is uneven with a heavier weight on minority class.

\section{Performance evaluation and simulation} \label{sec:sim}

Before we present the results of our simulation studies, we describe the performance measures used for classification problem, emphasizing that the G-mean measure is most aptly for imbalanced datasets.

\subsection{Performance metrics}

For classification problems, the most common performance statistics is accuracy, which is the proportion of all observations that were correctly classified. For obvious reasons, this is an irrelevant measure for imbalanced datasets. Let us consider three other performance statistics: Recall, Precision, and F1-score. We can compute performance statistics for each class, $i = 1,2,\ldots,K$, and aggregate them with an average. The Recall statistics for class $i$, $R_i$, is defined to be the proportion of observations in class $i$ correctly classified. The Recall statistics is also sometimes called the sensitivity. The Precision statistics, $P_i$, is the proportion predictions in class $i$ that were correctly classified. The F1-score, $\text{F1}_i$, is the harmonic average of Recall and Precision and is therefore equal to $2 \cdot (R_i \times P_i)/(R_i + P_i)$.

Consider the hypothetical example of three insurance claim frequencies for classes. Note that this illustrates highly imbalanced observations. Table \ref{tab:confusion} is a confusion matrix that is often used to describe the performance of a classifier. The columns refer to the actual observations and the rows refer to the predicted ones. Suppose the classes are designated respectively as $1,2,3$, with ``Claim 0'' as class 1, ``Claim 1'' as class 2, and ``Class $2^+$'' as class 3.  For ``Claim $2^+$'', we can compute the Recall $R_3 = 20/20 = 1.00$, Precision $P_3 = 20/1,020 = 0.02$, and F1-score $\text{F1}_3 = 2*(1*0.02)/(1+0.02) = 0.04$.

\begin{table}[H]
\caption{Confusion matrix of a hypothetical example}
\centering
\begin{tabular}{ccr|r|r|r}
	& & \multicolumn{3}{c}{Actual} & \\ \cline{3-5}
	 & \multicolumn{1}{c|}{} & Claim 0&Claim 1&Claim $2^+$ & Total rows\\ \cline{2-6} 
	\multirow{3}{*}{\rotatebox[origin=c]{90}{Predicted}} & \multicolumn{1}{|l|}{Claim 0}&9,000&0&0 & \multicolumn{1}{r|}{9,000}\\ \cline{2-6} 
	&\multicolumn{1}{|l|}{Claim 1}&0&500&0 & \multicolumn{1}{r|}{500} \\ \cline{2-6} 
	&\multicolumn{1}{|l|}{Claim $2^+$}&1,000&0&20 & \multicolumn{1}{r|}{1,020} \\ \cline{2-6}
	& \multicolumn{1}{l|}{Total columns} & 10,000& 500 & 20 &  \multicolumn{1}{r|}{10,520} \\ \cline{3-6}
\end{tabular}
\label{tab:confusion}
\end{table}

When we aggregate the results to provide a single measure of performance for a given classifier, we define statistics such as Macro Precision and Macro Recall to be the arithmetic average of the respective performance statistics for each class.  We define Macro F1-score as the reciprocal of the harmonic average of Macro Precision and Macro Recall, that is, $\text{Macro F1-score} = (2*\text{Macro Precision}*\text{Macro Recall})/(\text{Macro Precision}+\text{Macro Recall})$. For the Recall statistics, we shall use the geometric average and define
\begin{equation}
\text{G-mean} = (R_1 \times R_2 \times \cdots \times R_K)^{1/K}.
\end{equation}
Finally,  when comparing classifiers, a better performing classifier is one that gives a larger value for each of these aggregate statistics. 

If we take the log of both sides of the G-mean statistics, we get an average of the log of all the Recall statistics. For log transformation, the result gives a balance of the importance of accurately classifying observations for all classes. For imbalanced datasets, this would mean that it is perfectly acceptable to sacrifice mis-classifications of a majority class to correctly classify more of the minority class. Indeed, it has been discussed that the Recall, or sensitivity, is usually a more interesting measure for imbalanced classification. For these reasons, the G-mean is therefore a reasonable performance measure for imbalanced datasets. The G-mean concept is credited to the work of \citet{fowlkes1983gmean}. Table \ref{tab:metrics} provides the detailed values of the performance metrics resulting from our hypothetical example.

\begin{table}[H]
\caption{Performance metrics of the hypothetical example in Table \ref{tab:confusion}}
\centering
\begin{tabular}{c P{4cm}|P{4cm}|P{4.4cm}|}
	& \multicolumn{3}{c}{Values of performance metrics} \\ \cline{2-4}
	\multicolumn{1}{c|}{$i$} & Recall ($R_i$) & Precision ($P_i$) & F1-score ($\text{F1}_i$) \\ \hline 
	\multicolumn{1}{|c|}{Claim 0}& $R_1 = 0.90$ & $P_1 = 1.00$ & $\text{F1}_1= 0.95$ \\ \hline 
	\multicolumn{1}{|c|}{Claim 1}& $R_2 = 1.00$ & $P_2 = 1.00$ & $\text{F1}_2= 1.00$ \\ \hline 
	\multicolumn{1}{|c|}{Claim $2^+$}& $R_3 = 1.00$ & $P_3 = 0.02$ & $\text{F1}_3= 0.04$ \\ \hline
	 \multicolumn{1}{c|}{} &   &   &   \\ [\dimexpr-\normalbaselineskip+4pt]
	\multicolumn{1}{c|}{} & G-mean  $\sqrt[3]{(0.9)(1)(1)} = 0.965$ & Macro Precision  $\frac{1}{3}(1+1+0.02) = 0.673$ & Macro F1-score $\frac{2*0.673*0.967}{0.673*0.967} = 0.79$ \\ [4ex] \cline{2-4}
\end{tabular}
\label{tab:metrics}
\end{table}

\subsection{Simulation} \label{sub:sim}

To investigate how \texttt{SAMME.C2} algorithm performs relative to other boosting algorithms, the first crucial step is to perform this test based on a simple simulated dataset. For comparative models, as already alluded, we chose four other boosting algorithms that are potentially viable to compete when handling imbalanced multi-class classification problems: (a) \texttt{SAMME}, (b) \texttt{SAMME} with \texttt{SMOTE}, (c) \texttt{RUSBoost}, and (d) \texttt{SMOTEBoost}. Each of these methods has been described in the previous section.

To generate a simulated dataset, we utilize the \textit{Scikit-learn} Python module described in \citet{pedregosa2011scikit}. The module is a user-friendly tool for applying ``state-of-the-art machine learning algorithms.'' We find the built-in function called \textpython{make\_classification} with parameterization executed as

\begin{scriptsize}
\begin{verbatim}
"""Make Simulation"""
from sklearn.datasets  import make_classification

X, y = make_classification(n_samples=100000, n_features=50, n_informative=5-, n_redundant=0, n_repeated=0,
n_classes=3, n_clusters_per_class=2, class_sep=2, flip_y=0, weights=[0.96,0.035,0.005], random_state=16)
\end{verbatim}
\end{scriptsize}

This generates 100,000 samples with 50 features, and 3 classes were generated, deliberately creating a highly imbalanced dataset by setting the ratio for each class as 96\%, 3.5\%, and 0.5\%, respectively. We ask the reader to refer to the package for an explanation of the other parameters. For training, we use 75\% of the data, and the rest was used for testing.

All boosting algorithms used for the analysis utilize one-depth-decision tree as a weak classifier. 200 weak classifiers are linearly combined with weights. To implement the cost values for \texttt{SAMME.C2}, we employed Genetic Algorithm (GA), which is a directed random search technique invented by \citet{holland1975} and described in \citet{muhlenbein1997GA}. This same genetic algorithm for implementing cost function will be used in the empirical application. See also \citet{bhowan2010genetic}. In the algorithm, after setting the population set, each individual in the population is tested with a fitness function. The best individuals are chosen and a new individual is made through cross-over. By adding some error to a new one, it is mutated, and new population is generated through this mutation. With new population, every step is repeated. To select the best cost vector, the G-mean value is designated as the fitness function.

Each boosting algorithm was then trained using the training data, and performance is evaluated on the training data. The results of the performance of the various models are summarized in Table \ref{tab:simresults}. According to these results, in terms of the G-mean,  \texttt{SAMME.C2}, with a G-mean of 0.93, outperforms all the other models, with \texttt{SAMME} with \texttt{SMOTE} not far behind at 0.90. However, an examination of each class Recall statistics reveals that  \texttt{SAMME.C2} outperforms all other models correctly classifying those belonging to the minority class. But clearly, this comes with the price of having the worst Recall statistics for the majority class, ``Claim 0.'' 

\begin{table}[htbp]
\caption{Comparing the performance measures using G-mean based on the simulated dataset}
\centering
\begin{tabular}{lP{2.5cm}|P{2.5cm}|P{2.5cm}|P{2.5cm}|P{2.5cm}|}
	& \multicolumn{5}{c}{Recall statistics} \\ \cline{2-6}
	\multicolumn{1}{P{1.5cm}|}{\hspace{1.5cm} Class} & \hspace{2cm} \texttt{SAMME} &\texttt{SAMME} with \texttt{SMOTE} &  \hspace{2cm} \texttt{RUSBoost} &  \hspace{2cm} \texttt{SMOTEBoost} &  \hspace{2cm} \texttt{SAMME.C2} \\ \hline 
	\multicolumn{1}{|l|}{Claim 0}& 1.00 & 0.96 & 0.88 & 0.99 & 0.88 \\ \hline 
	\multicolumn{1}{|l|}{Claim 1}& 0.84 & 0.93 & 0.88 & 0.91 & 0.95 \\ \hline 
	\multicolumn{1}{|l|}{Claim $2^+$}& 0.42 & 0.83 & 0.30 & 0.76 & 0.95 \\ \hline
	\multicolumn{1}{l|}{G-mean} & 0.71  & 0.90 & 0.62 & 0.88 & 0.93 \\   \cline{2-6}
\end{tabular}
\label{tab:simresults}
\end{table}

\section{Empirical data: telematics} \label{sec:empirical}

The goal of this paper is to analyze and understand driver's behavior using telematics. In this section, we demonstrate the strength of our \texttt{SAMME.C2} algorithm as a predictive model for accident frequencies, the number of times a driver had a claim during the observation period. Using this as a predictive model, we subsequently analyze the effects of the value added by the telematics information. Insurance companies have long been using traditional variables for driver's risk classification, such as age and gender. The advancement of technology have led insurers to offer innovative product such as UBI that uses telematics to better classify and price risks with the understanding that such additional information provides better tools for learning driver behavior.

\subsection{Description of data} \label{sub:data}

Our telematics dataset has been acquired from a Canadian-owned co-operative that offers insurance and investment products; its UBI program was launched in Ontario in year 2013. Raw data consisted of information from drivers who participated in this program and were observed during the period of 2013-2016, with nearly 100,000 records. The response variable of interest is accident frequency, the number of accidents observed per driver, and we can observe 0 (no claims), 1 (exactly one claim), or 2 (two or more claims). For training, we have a total  of 50,301 observations; 48,822 have no claims, 1,430 have exactly one claim, and only 49 have 2 or more claims. We see the presence of highly imbalanced classes: 97.1\% with no claims, 2.8\% with exactly one claim, and only 0.1\% with two or more claims. We have no missing values of claims frequency in the dataset.

\begin{figure}[htbp]
\centering
\includegraphics[scale=0.85]{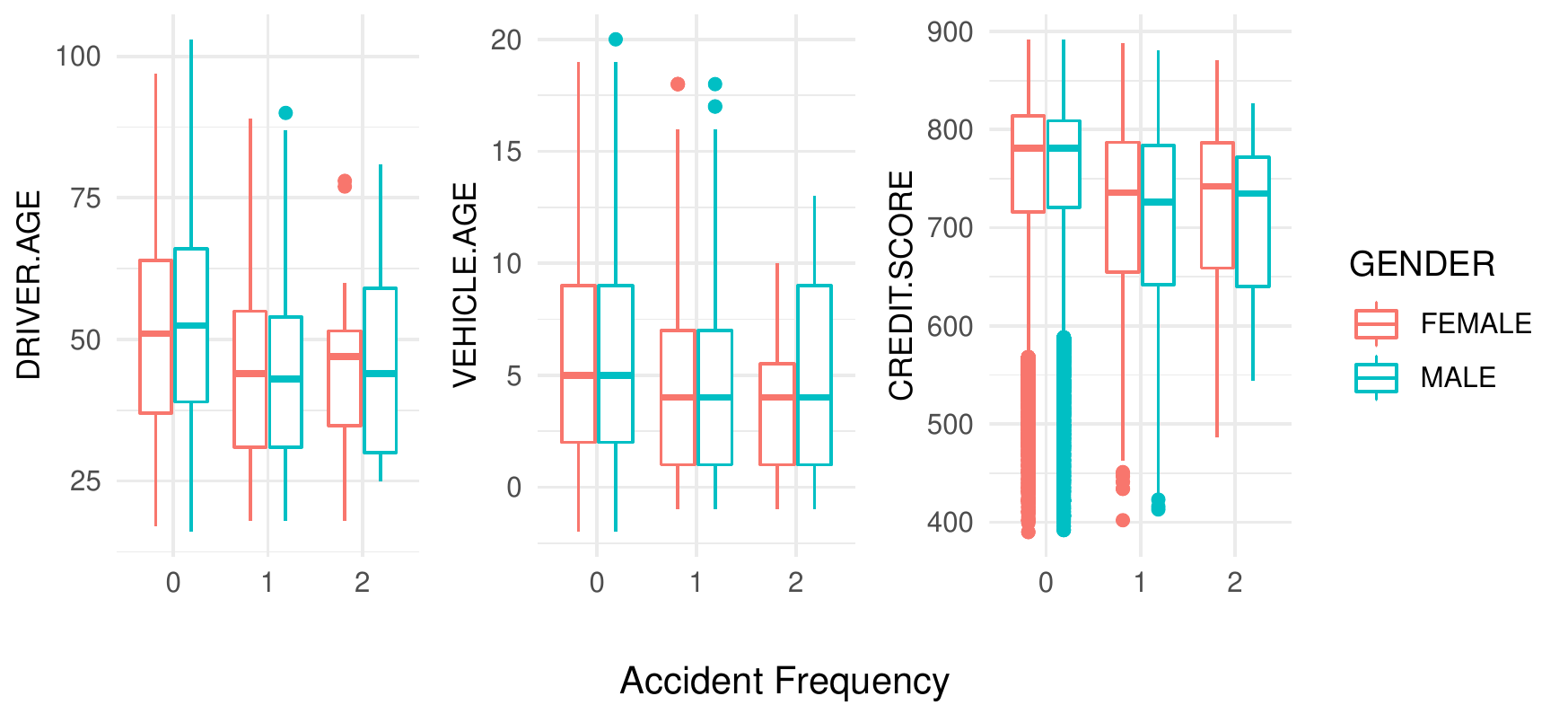}
\caption{Analysis of claims frequency by gender} \label{fig2:gender}
\end{figure}

We have a total of 49 potential predictor variables, broken down into 10 traditional (e.g., driver age, gender) and the rest are telematics driven predictor variables. The descriptions for each variable in our dataset is summarized in Table \ref{tab:VD}. To have a preliminary understanding of the data, we present in Table \ref{tab:summ1} summary statistics of three traditional continuous variables: DRIVER.AGE, VEHICLE.AGE, and CREDIT.SCORE. For example, we see that average driver age is 51.3, with 16.0 and 103.0 as the youngest and oldest drivers, respectively, in our records. The cohort of drivers in our dataset has more mature driving experience that falls within the middle age groups. According to Figure \ref{fig2:gender}, there does not seem to be significant difference between male and female with respect to accident frequency even after controlling for the effect of DRIVER.AGE, VEHICLE.AGE, or CREDIT.SCORE.

\begin{figure}[htbp]
\centering
\includegraphics[scale=0.8]{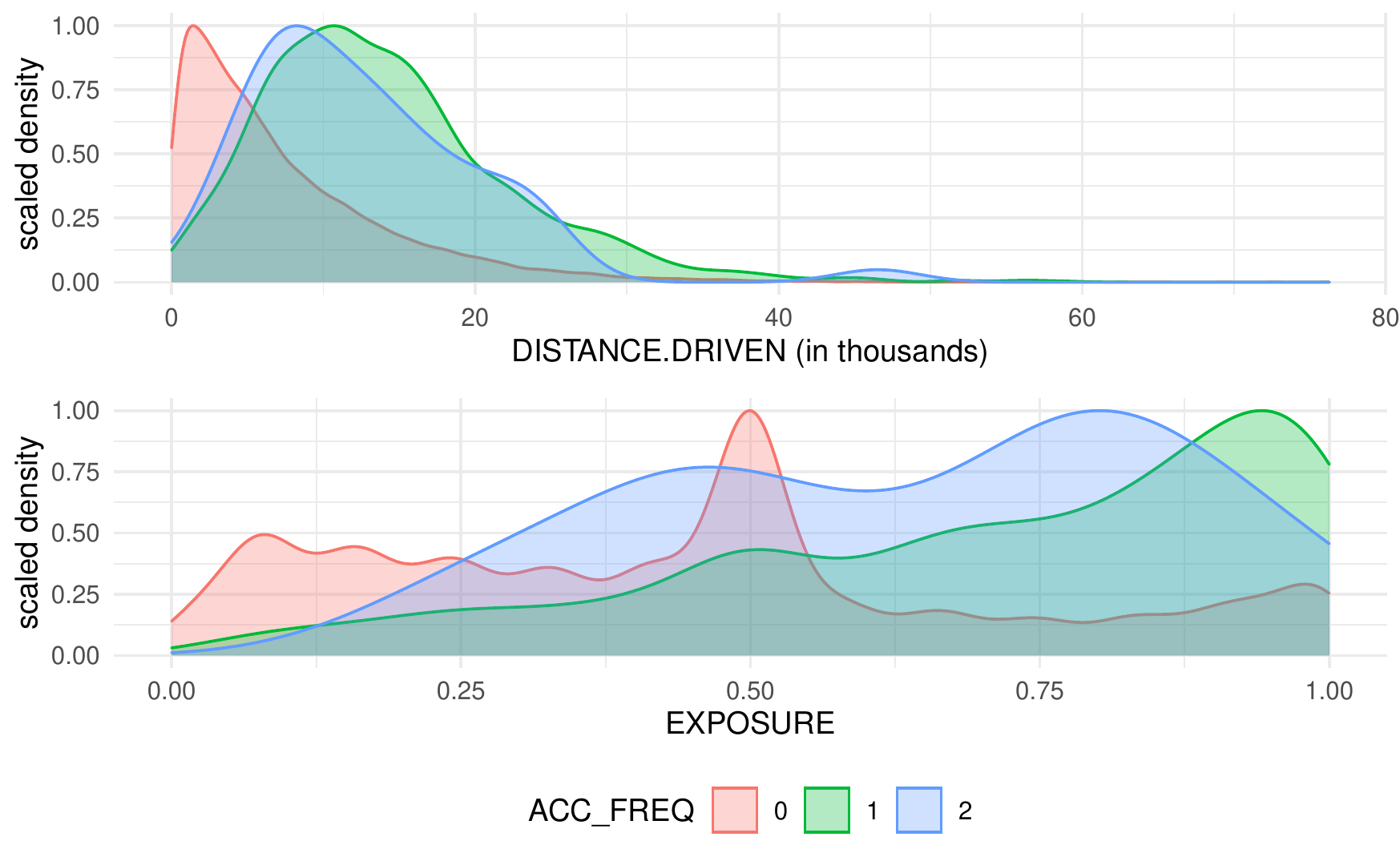}
\caption{Analysis of claims frequency by Exposure and Distance Driven} \label{fig3:telem}
\end{figure}

Table \ref{tab:summ2} provides summary statistics of two telematics variables: DISTANCE.DRIVEN and EXPOSURE . For example, the average distance traveled is 7,555.3 kms for those without claims, 14,155.4 for those with exactly one claim, and 12,834.89 for those with two or more claims. Broadly speaking, this appears to indicate that the more distance traveled, the more likely there will be at least one claim. Figure \ref{fig3:telem} also seems to confirm that relatively compares the shape of the distribution of distance driven and exposure according to accident frequency.

\begin{table}[htbp]
\caption{Summary statistics of three (continuous) traditional variables}
\centering
\begin{tabular}{l|rrrrrrr}
	\cline{2-8}
	& Mean & Std Dev & Min & Q1 & Median & Q3 & Max \\ \hline
	DRIVER.AGE & 51.3 & 16.8 & 16.0 & 38.0 & 51.0 & 65.0 & 103.0 \\ \hline
	VEHICLE.AGE & 5.7 & 4.49 & -2.0 & 2.0 & 5.0 & 9.0 & 20.0 \\ \hline
	CREDIT.SCORE & 754.8 & 88.0 & 390.0 & 716.0 & 780.0 & 811.0 & 892.0 \\ \hline
\end{tabular}
\label{tab:summ1}
\end{table}

\begin{table}[htbp]
\caption{Summary statistics of two telematics variables}
\centering
\resizebox{!}{1.6cm}{
\begin{tabular}{l|ccrrrrrrr}
	\hline
	Variable & Acc Freq & Count & Mean & Std Dev & Min & Q1 & Median & Q3 & Max \\ \hline
	DISTANCE.DRIVEN & 0 & 48822	& 7555.3 & 7149.4 & 0.1 & 2374.8 & 5395.7 & 10592.7 & 76271.8 \\ \cline{2-10}
	 & 1 & 1430 & 14155.4 & 8257.3 & 253.9 & 8319.7 & 12657.4 & 18161.2 & 58759.2 \\ \cline{2-10}
	 & 2 & 49 & 12834.9 & 7925.9 & 2247.8 & 7408.1 & 11408.3 & 16621.3 & 46527.4 \\ \hline
	EXPOSURE & 0 & 48822 & 0.49 & 0.31 & 0.00 & 0.24 & 0.50 & 0.73 & 1.08 \\ \cline{2-10}
	 & 1 & 1430 & 0.78 & 0.25 & 0.02 & 0.64 & 0.89 & 1.00 & 1.06 \\ \cline{2-10}
	 & 2 & 49 & 0.74 & 0.26 & 0.23 & 0.50 & 0.80 & 1.00 & 1.06\\ \hline
\end{tabular}}
\label{tab:summ2}
\end{table}

We can broadly classify the telematics variables into those that are considered driving maneuvers (braking, accelerations, and harsh events that include left or right turn events) and those that do not fall into this category. Most variables falling in the latter category relate to how much time drivers are on the road (percentage of driving spent during each day of the week, exposure, and distance driven). According to Table \ref{tab:VD}, there are essentially 4 types of driving maneuvers: braking, accelerations, left turn events, and right turn events.

\begin{figure}[htbp]
\centering
\includegraphics[scale=0.75]{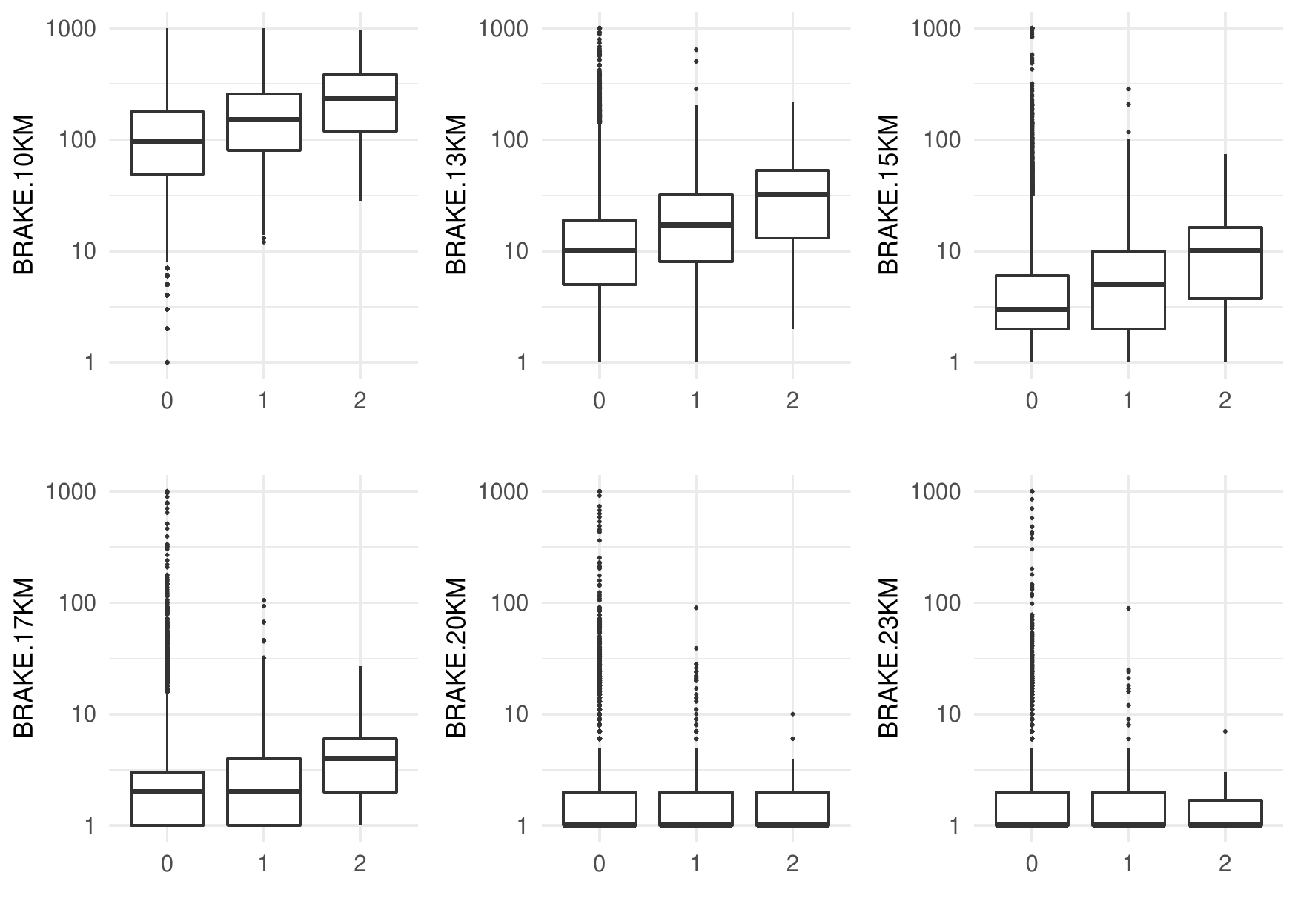}
\caption{Analysis of claims frequency by Exposure and Distance Driven} \label{fig4:brakes}
\end{figure}

BRAKE.xxKM refers to the number of sudden brakes applied at different measures of deceleration: 10000/13000/15000/17000/20000/23000 km per h/s. The more rapid the deceleration, the faster the car is being driven when a sudden brake is applied. Figure \ref{fig4:brakes} displays a boxplot of the frequency of these sudden brakes according to accident frequency. The $y$-axis has been measured on a logarithmic scale. First, note that the application of brakes at higher deceleration is less frequently observed. Second, the more frequent application of brakes for different decelerations shows the increase in likelihood of an accident. 

\subsection{Performance evaluation} \label{sub:perf}

Using a test dataset, we evaluate the performance of \texttt{SAMME.C2} against other classification models as earlier described. For this test dataset, we have a total of 21,574 observations: 20,901 have no claims, 650 have exactly one claim, and only 23 have two or more claims. Every settings for the various algorithms are identical to the ones described in Section \ref{sec:sim} on simulation. For the cost vector for \texttt{SAMME.C2}, this has been preprocessed using genetic algorithm for which the assigned cost is unique to each class in each observation.

The results summarized in Table \ref{tab:results} are rather very encouraging for \texttt{SAMME.C2}. This is because in terms of the G-mean performance metrics, which is 0.71 for this algorithm, it far outperforms all the other models. Although \texttt{SAMME.C2} made some sacrifices for accuracy prediction of the majority class (here, the case of zero claims), it produced a superior outcome of predicting the other classes considered minority. Several can be further deduced from these results. As in the simulation studies, the worst classifiers are either those that do not resample (\texttt{SAMME}) or do but uses only undersampling (\texttt{RUSBoost}). In the simulation studies, \texttt{SAMME.C2} only slightly outperformed the other two boosting algorithms that uses the \texttt{SMOTE} resampling method. However, when real data is used, our results demonstrate that the use of cost-sensitive learning mechanism can outperform mechanisms that use resampling methods.

\begin{table}[H]
\caption{Comparing the performance measures using G-mean based on the telematics dataset}
\centering
\begin{tabular}{lP{2.4cm}|P{2.3cm}|P{2.3cm}|P{2.3cm}|P{2.3cm}|}
	& \multicolumn{5}{c}{Recall statistics} \\ \cline{2-6}
	\multicolumn{1}{P{2.1cm}|}{Accident Frequency} & \hspace{2cm} \texttt{SAMME} &\texttt{SAMME} with \texttt{SMOTE} &  \hspace{2cm} \texttt{RUSBoost} &  \hspace{2cm} \texttt{SMOTEBoost} &  \hspace{2cm} \texttt{SAMME.C2} \\ \hline 
	\multicolumn{1}{|l|}{Accident 0}& 1.00 & 0.84 & 0.68 & 0.99 & 0.70 \\ \hline 
	\multicolumn{1}{|l|}{Accident 1}& 0.06 & 0.38 & 0.46 & 0.18 & 0.65 \\ \hline 
	\multicolumn{1}{|l|}{Accident $2^+$}& 0.00 & 0.35 & 0.04 & 0.13 & 0.78  \\ \hline
	\multicolumn{1}{l|}{G-mean} & 0.02  & 0.48 & 0.24 & 0.28 & 0.71 \\   \cline{2-6}
\end{tabular}
\label{tab:results}
\end{table}

\subsection{Interpretation} \label{sub:effect}

Several researchers and industry practitioners continue to be interested in the usefulness of vehicle telematics information to build more customized pricing models. In this subsection, we show how the results of using \texttt{SAMME.C2} algorithm helps enhance our understanding of driving behavior. 

\subsubsection{Traditional vs telematics predictors} \label{sub:telem}

To evaluate the added value of telematics information, we compare the results of using \texttt{SAMME.C2} with the 10 traditional variables only and with all 49 traditional and telematics variables. The results are summarized in Table \ref{tab:trad}.

From the confusion tables, it can easily be calculated that for the \texttt{SAMME.C2} with all 49 variables, the resulting G-mean is 0.71; when only traditional variables are used, the resulting G-mean is 0.58. Hence, this suggests that prediction models from \texttt{SAMME.C2} that includes telematics information outperform those with just traditional predictor variables. Prior studies have similarly concluded that claims frequency models that contain telematics information are far better than just those with classical predictor variables. See \citet{boucher2017gam}, \citet{ayuso2019auto}, and \citet{guillen2020telem}. Note that in this table, the diagonal values give the numbers of correctly classified instances for each accident frequency according to the \texttt{SAMME.C2} algorithm. 

\begin{table}[htbp]
\caption{Confusion tables based on the model fit of \texttt{SAMME.C2}} \label{tab:trad}
\centering
\renewcommand\arraystretch{1.3}
\begin{minipage}[t]{.48\textwidth}
{\par\centering With traditional variables only\par}
\resizebox{!}{1.9cm}{
\begin{tabular}{ccr|r|r|r}
	& & \multicolumn{3}{c}{Actual} & \\ \cline{3-5}
	 & \multicolumn{1}{c|}{} & Acc 0&Acc 1&Acc $2^+$ & Tot row \\ \cline{2-6} 
	\multirow{3}{*}{\rotatebox[origin=c]{90}{Predicted}} & \multicolumn{1}{|l|}{Acc 0}&11674 & 175 & 1 &  \multicolumn{1}{r|}{11850}\\ \cline{2-6} 
	&\multicolumn{1}{|l|}{Acc 1}&5319&274&3 & \multicolumn{1}{r|}{5596} \\ \cline{2-6} 
	&\multicolumn{1}{|l|}{Acc $2^+$}&3908&201&19 & \multicolumn{1}{r|}{4128} \\ \cline{2-6}
	& \multicolumn{1}{l|}{Tot col} & 20901& 650 & 23 &  \multicolumn{1}{r|}{21574} \\ \cline{3-6}
\end{tabular}}
\end{minipage}
\hfill
\begin{minipage}[t]{.48\textwidth}
{\par\centering With traditional and telematics variables\par}
\resizebox{!}{1.9cm}{
\begin{tabular}{ccr|r|r|r}
	& & \multicolumn{3}{c}{Actual} & \\ \cline{3-5}
	 & \multicolumn{1}{c|}{} & Acc 0&Acc 1&Acc $2^+$ & Tot row \\ \cline{2-6} 
	\multirow{3}{*}{\rotatebox[origin=c]{90}{Predicted}} & \multicolumn{1}{|l|}{Acc 0}&14553&108&0 & \multicolumn{1}{r|}{14661}\\ \cline{2-6} 
	&\multicolumn{1}{|l|}{Acc 1}&5669&420&5 & \multicolumn{1}{r|}{6049} \\ \cline{2-6} 
	&\multicolumn{1}{|l|}{Acc $2^+$}&679&122&18 & \multicolumn{1}{r|}{819} \\ \cline{2-6}
	& \multicolumn{1}{l|}{Tot col} & 20901 & 650 & 23 &  \multicolumn{1}{r|}{21574} \\ \cline{3-6}
\end{tabular}}
\end{minipage}
\end{table}

\begin{figure}[htbp]
\centering
\includegraphics[width=\linewidth]{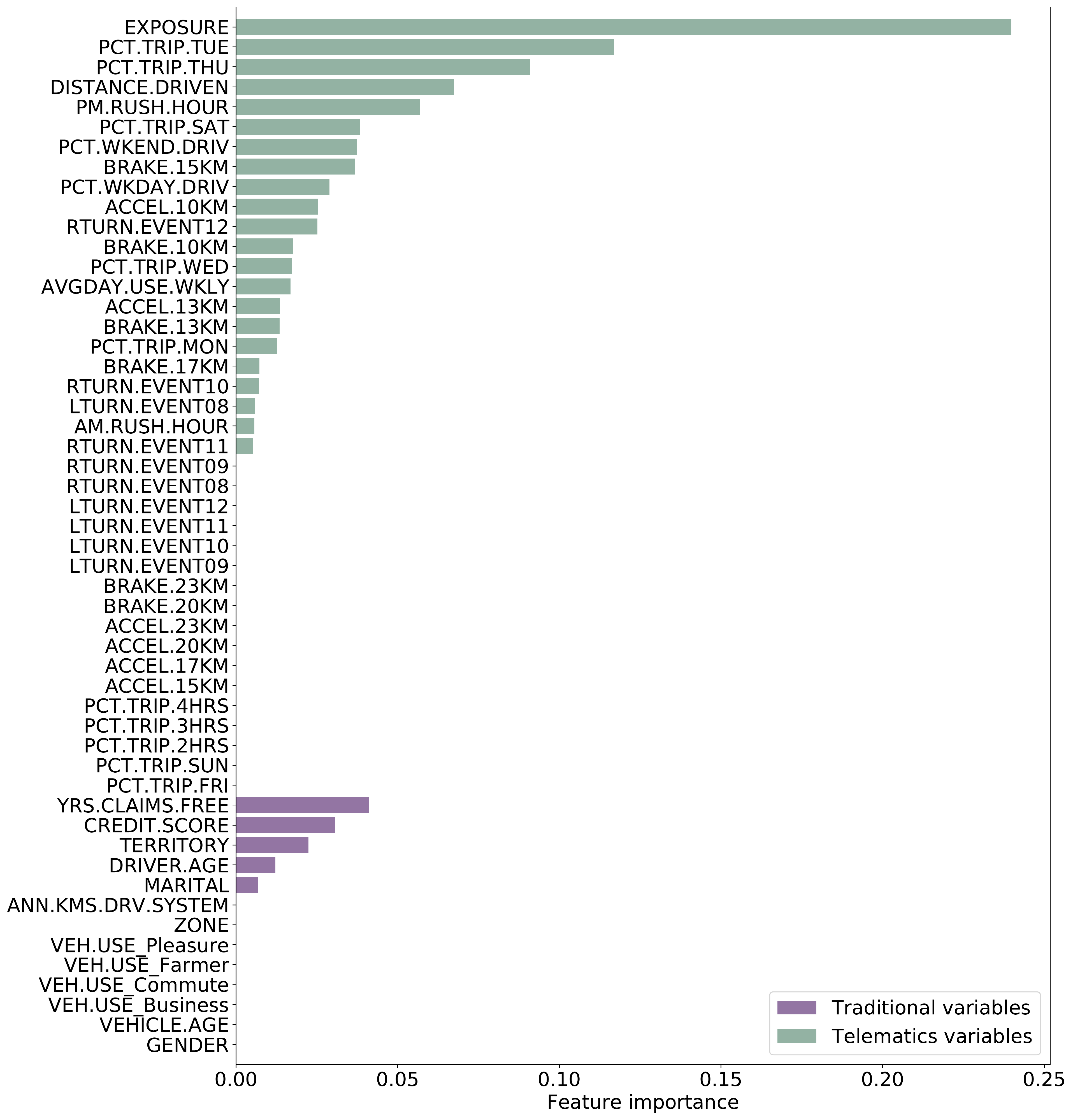}
\caption{Feature importance based on the \texttt{SAMME.C2} predictions - using training data} \label{fig5:featimp}
\end{figure}

Figure \ref{fig5:featimp} depicts the relative importance of the feature variables, with the telematics variables listed first and followed by the traditional variables. There are quite a number of conclusive evidence that we can draw and consider from this decomposition of the contributions of the feature variables. First, this figure highlights that there are many more telematics than traditional variables that are better predictor variables of accident frequency. For example, the top five telematics features (EXPOSURE, PCT.TRIP.TUE, PCT.TRIP.THU, DISTANCE.DRIVEN, and PM.RUSH.HOUR) are relatively more important than the top traditional feature (YRS.CLAIMS.FREE). This seems to suggest that traditional features that are usually considered significant predictors may be replaced by information drawn from telematics. Second, in terms of feature importance, being the top feature variable, EXPOSURE far outweighs all other variables. This variable is defined to be the frequency of being on the road driving within a 24-hour cycle.

Third, as described in Section \ref{sub:data}, we may classify the telematics variables into those considered driving maneuvers and those that relate to how much time drivers are on the road. The feature importance suggests that driving maneuvers are less important predictor variables than the time and distance travelled on the road. For example, the top six telematics variables (EXPOSURE, PCT.TRIP.TUE, PCT.TRIP.THU, DISTANCE.DRIVEN, and PM.RUSH.HOUR) are not at all related to driving maneuvers. In some sense, this can be explained by the possible effect of a change in driver behavior knowing that there is constant monitoring when a telematics device is installed. These driving maneuvers are more indicative of this change than the frequency and distance the driver will spend on the road. Finally, we find that gender has little effect in predicting accident frequency. This is in line with previous studies, such as in \citet{ayuso2016gender}, that gender discrimination may not be necessary with telematics information. The difference in driving behavior according to gender may well be reflected in the telematics information. However, we also note that this might simply be what our data suggest, that irrespective of the models used, there is no significant difference by gender. We found this observation in our preliminary data investigation. See Figure \ref{fig3:telem}.

\subsubsection{Further understanding effects of telematics} \label{sub:effects}

To understand the predictive power of \texttt{SAMME.C2}, we considered drawing three sampled observations from our dataset. We selected these observations according to a common set of traditional variables: a single, male, 48-year-old driver with the range of a high credit score and have more than 18 claims-free years. For all three observations, all the values of the telematics variables have been varied. For example, the observations have respective EXPOSURE of 0.65, 0.50, and 0.14, and respective DISTANCE.DRIVEN of 23802, 7717, and 3340. Note that the first and last observations did not have any accidents, but the second one had two or more accidents. Using the model with all traditional and telematics variables, all three observations were correctly classified, but if only traditional variables were used, all three would have been incorrectly classified. These details are provided in Table \ref{tab:sample}.

\begin{table}[htbp]
\caption{Comparing predicted probabilities and classification based on the model fit of \texttt{SAMME.C2} for three observations}  \label{tab:sample}
\centering
\resizebox{!}{2.7cm}{
\begin{tabular}{|c|c|c|c|c|c|c|c|c|c|c|}
	\hline
	\multirow{8}{*}{\rotatebox[origin=r]{270}{DRIVER.AGE}} & \multirow{8}{*}{\rotatebox[origin=r]{270}{MARITAL}} & \multirow{8}{*}{\rotatebox[origin=r]{270}{CREDIT.SCORE}} & \multirow{8}{*}{\rotatebox[origin=r]{270}{VEHICLE.AGE}} & \multirow{8}{*}{\rotatebox[origin=r]{270}{GENDER}} & \multirow{8}{*}{\rotatebox[origin=r]{270}{YRS.CLMS.FREE}} & \multirow{8}{*}{\rotatebox[origin=r]{270}{EXPOSURE}} & \multirow{8}{*}{\rotatebox[origin=r]{270}{DISTANCE.DRIVEN}} & \multirow{8}{*}{\parbox{2.2cm}{\centering Traditional plus Telematics: Predicted Probability (Predicted ACC\_FREQ)}} & \multirow{8}{*}{\parbox{2.2cm}{\centering Traditional Only:  Predicted Probability (Predicted ACC\_FREQ)}} & \multirow{8}{*}{\parbox{2.2cm}{\centering Actual ACC\_FREQ}} \\
	& & & & & & & & & & \\
	& & & & & & & & & & \\
	& & & & & & & & & & \\
	& & & & & & & & & & \\
	& & & & & & & & & & \\
	& & & & & & & & & & \\
	& & & & & & & & & & \\ \hline
	\multirow{3}{*}{48} & \multirow{3}{*}{Single} & \multirow{3}{*}{793--860} &  \multirow{3}{*}{0--2} & \multirow{3}{*}{Male} & \multirow{3}{*}{$\ge18$} & 0.65 & 23802 & 0.34 (0) & 0.34 (1) & 0 \\ \cline{7-11}
	& & & & & & 0.50 & 7717 & 0.35 ($2^+$)  & 0.34 (1) & $2^+$ \\ \cline{7-11}
	& & & & & & 0.14 & 3340 & 0.35 (0) & 0.34 (1) & 0 \\ \hline
\end{tabular}}
\end{table}

To further demonstrate the predictive power of the \texttt{SAMME.C2} with telematics, we consider three hypothetical drivers with common values of traditional variables: DRIVER.AGE=40, VEHICLE.AGE=10, GENDER=Female, MARITAL=Married, VEHICLE.USE=Commute, and ZONE=Rural. Table \ref{tab:hypo} provides the predicted classification varying the values of each of the 13 top ranked telematics variables. The values of these telematics variables have been carefully picked so that we can show a prediction of each class of ACC\_FREQ. Furthermore, we also preserve the observed relationships of these telematics variables to that previously discussed. For example, consider the hypothetical driver H1 for which we predict that this driver will have no accidents. Apart from the listed values of the common traditional variables, this H1 driver has EXPOSURE of 0.30, PCT.TRIP.TUE of 0.10, PCT.TRIP.THU of 0.10, DISTANCE.DRIVEN of 5000, and the rest as listed in the table. 

\begin{table}[htbp]
\caption{Predicted classification based on the model fit of \texttt{SAMME.C2} for hypothetical drivers varying telematics information but keeping same values of traditional variables: DRIVER.AGE=40, VEHICLE.AGE=10, GENDER=Female, MARITAL=Married, VEHICLE.USE=Commute, and ZONE=Rural} \label{tab:hypo}
\centering
\resizebox{!}{2.95cm}{
\begin{tabular}{|c|c|c|c|c|c|c|c|c|c|c|c|c|c|c|}
	\hline
	 \multirow{10}{*}{\parbox{2.2cm}{\centering Hypothetical driver}} & \multirow{10}{*}{\rotatebox[origin=r]{270}{EXPOSURE}} & \multirow{10}{*}{\rotatebox[origin=r]{270}{PCT.TRIP.TUE}} & \multirow{10}{*}{\rotatebox[origin=r]{270}{PCT.TRIP.THU}} & \multirow{10}{*}{\rotatebox[origin=r]{270}{DISTANCE.DRIVEN}} & \multirow{10}{*}{\rotatebox[origin=r]{270}{PM.RUSH.HOUR}} & \multirow{10}{*}{\rotatebox[origin=r]{270}{YRS.CLMS.FREE}} & \multirow{10}{*}{\rotatebox[origin=r]{270}{PCT.TRIP.SAT}} & \multirow{10}{*}{\rotatebox[origin=r]{270}{PCTWKEND.DRIV}} & \multirow{10}{*}{\rotatebox[origin=r]{270}{BRAKE.15KM}} & \multirow{10}{*}{\rotatebox[origin=r]{270}{CRED.SCORE}} & \multirow{10}{*}{\rotatebox[origin=r]{270}{PCTWKDAY.DRIV}} & \multirow{10}{*}{\rotatebox[origin=r]{270}{ACCEL.10KM}} & \multirow{10}{*}{\rotatebox[origin=r]{270}{RTURN.EVENT12}} & \multirow{10}{*}{\parbox{2.3cm}{\centering Predicted ACC\_FREQ}} \\
	& & & & & & & & & & & & & & \\
	& & & & & & & & & & & & & & \\
	& & & & & & & & & & & & & & \\
	& & & & & & & & & & & & & & \\
	& & & & & & & & & & & & & & \\
	& & & & & & & & & & & & & & \\
	& & & & & & & & & & & & & & \\
	& & & & & & & & & & & & & & \\
	& & & & & & & & & & & & & & \\ \hline
	H1 & 0.3 & 0.10 & 0.10 & 5000 & 0.10 & 30 & 0.10 & 0.50 & 10 & 850 & 0.50 & 30 & 10 & 0 \\ \hline
	H2 & 0.8 & 0.13 & 0.13 & 15000 & 0.15 & 15 & 0.20 & 0.20 & 20 & 650 & 0.80 & 70 & 20 & 1 \\ \hline
	H3 & 0.8 & 0.18 & 0.18 & 15000 & 0.20 & 10 & 0.08 & 0.80 & 150 & 650 & 0.20 & 150 & 150 & $2^+$ \\ \hline
\end{tabular}}
\end{table}

\begin{figure}[htbp]
\centering
\includegraphics[width=\linewidth]{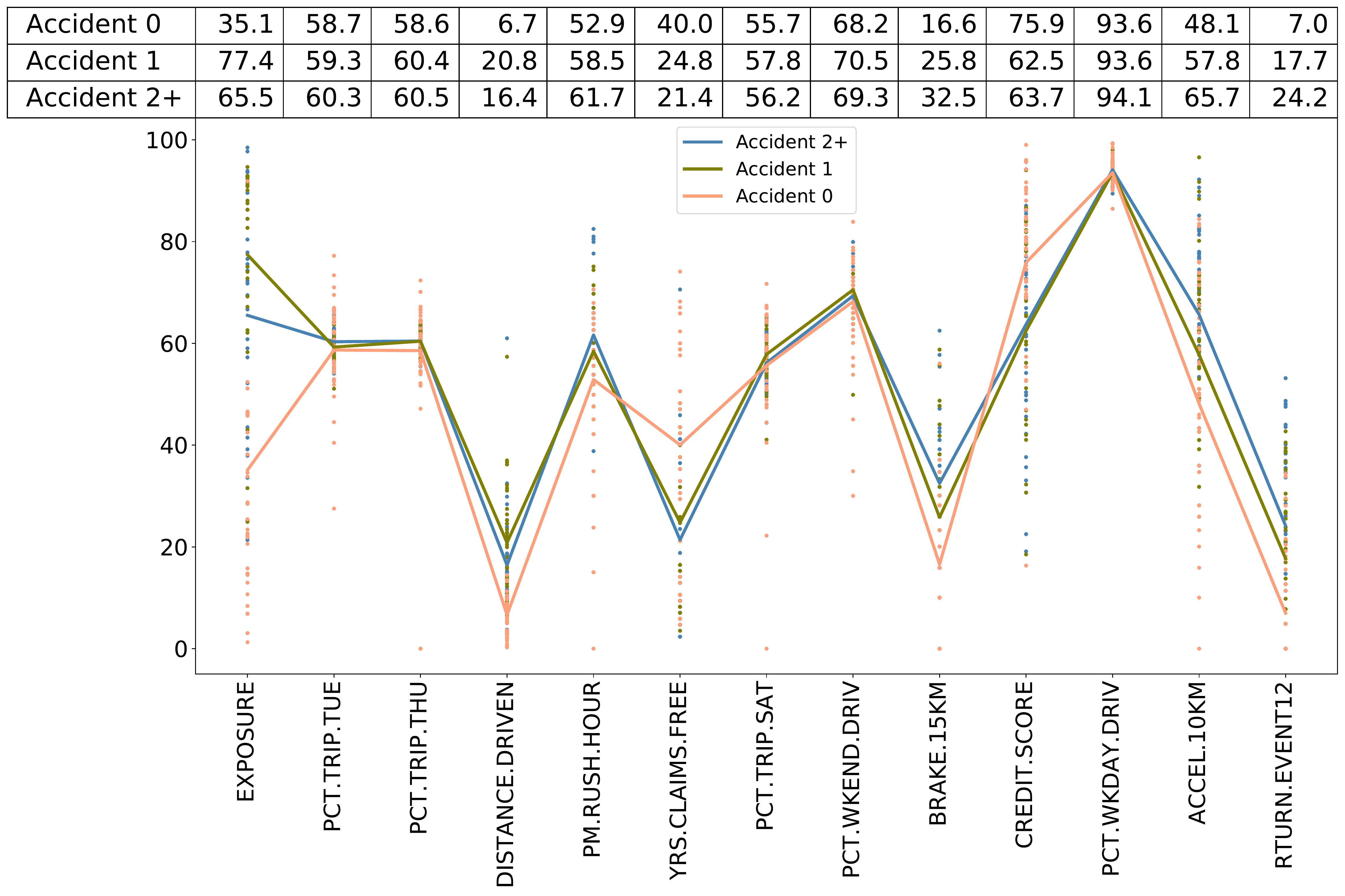}
\caption{Distribution and mean values of the 13 top ranked variables for a sample of 120 drivers} \label{fig6:people}
\end{figure}

To further understand the effects of the 13 top ranked variables, we display Figure \ref{fig6:people}, which provides the resulting distributions of these variables with mean point emphasized and distinguished according to accident frequency. Each dot represents a sampled observation of 120 drivers drawn from our training dataset. Because of the apparent differences in values for all 13 top ranked variables, we standardized each variable using the min-max normalization. This transformation leads us to a common measure of values that ranges between 0 and 100 for all variables. The smallest value is assigned a 0, the largest is assigned a 100, and all others are values anywhere in between. Notice also that of these 13 top ranked feature variables, only YRS.CLAIMS.FREE and CREDIT.SCORE are traditional variables. All the rest are telematics variables.

In order to understand this figure, consider the extremely rare case of having two or more accidents. According to the figure, when compared to accident frequencies of 0 or 1, there will be no distinction with respect to the following feature variables: PCT.TRIP.TUE, PCT.TRIP.THU, PCT.TRIP.SAT, and PCT.WKEND.DRIV. This is not at all surprising because of possible high correlation of these variables to the top feature variable: EXPOSURE. Recall that EXPOSURE is the average percentage of time spent driving within a 24-hour cycle; the higher the EXPOSURE, the higher the percentage of time spent driving each day of the week or on weekend. On the other hand, on an average basis, this same extremely rare case of having two or more accidents tends to exhibit: (a) higher EXPOSURE, (b) more DISTANCE.DRIVEN, (c) more frequent driving during PM.RUSH.HOUR, (d) fewer YRS.CLAIMS.FREE, (e) more frequent BRAKE.15KM, (f) lower CREDIT.SCORE, (g) higher ACCEL.10KM, and (h) more frequent RTURN.EVENT12.

This figure also shows the directional effect of the relationships of accident frequency to each of the 13 top ranked variable, if at all there is difference. It is also interesting to note that when it comes to EXPOSURE variable, it is true that those with zero accidents have lower EXPOSURE. The pattern between one accident and two or more accidents is not as intuitive because those with accidents of two or more tend to have generally lower EXPOSURE. We attribute this to the possibility that we only have very few observations with two or more accidents.

\begin{figure}[htbp]
 \includegraphics[width=.32\linewidth]{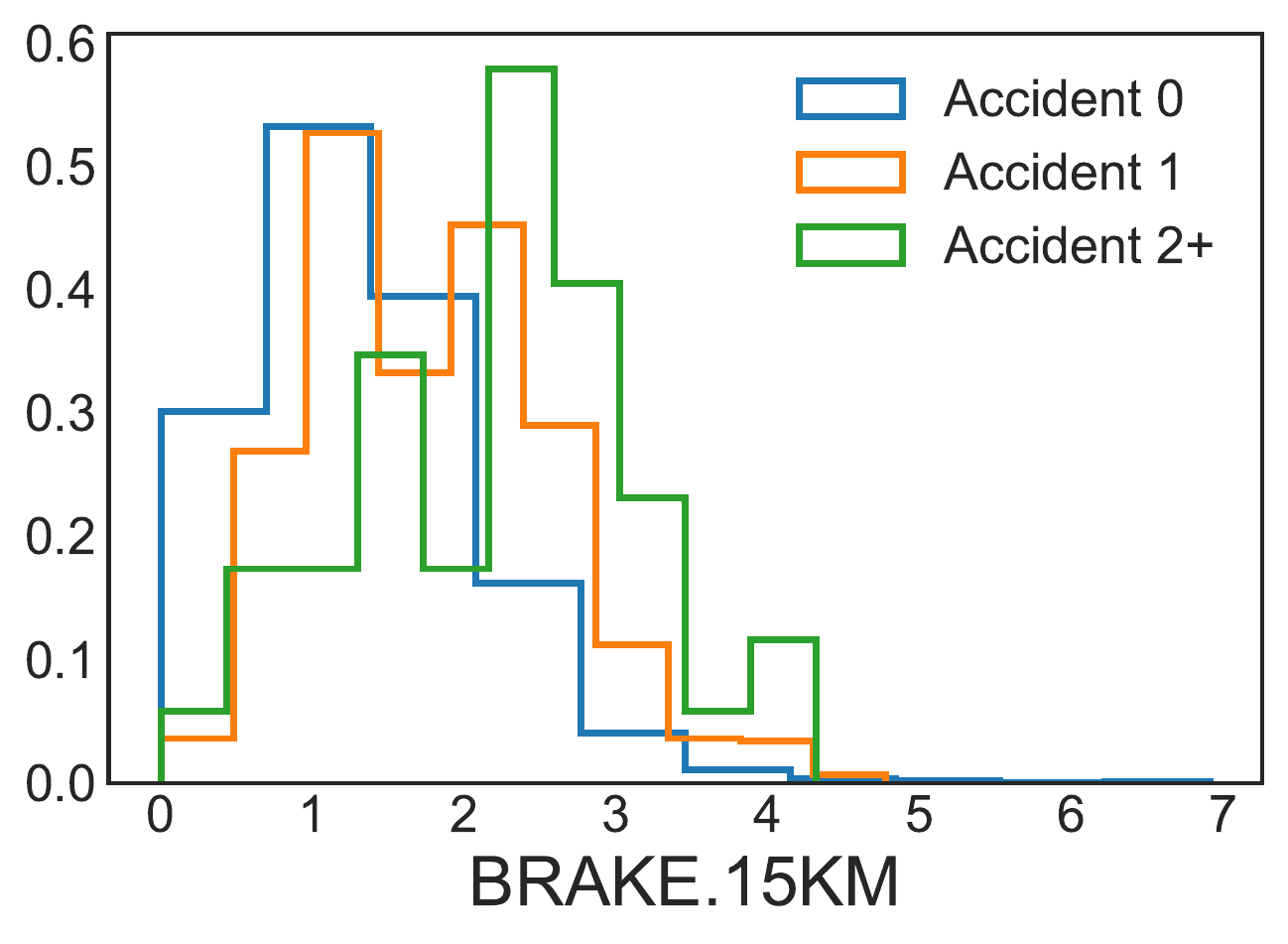} \hfill
 \includegraphics[width=.32\linewidth]{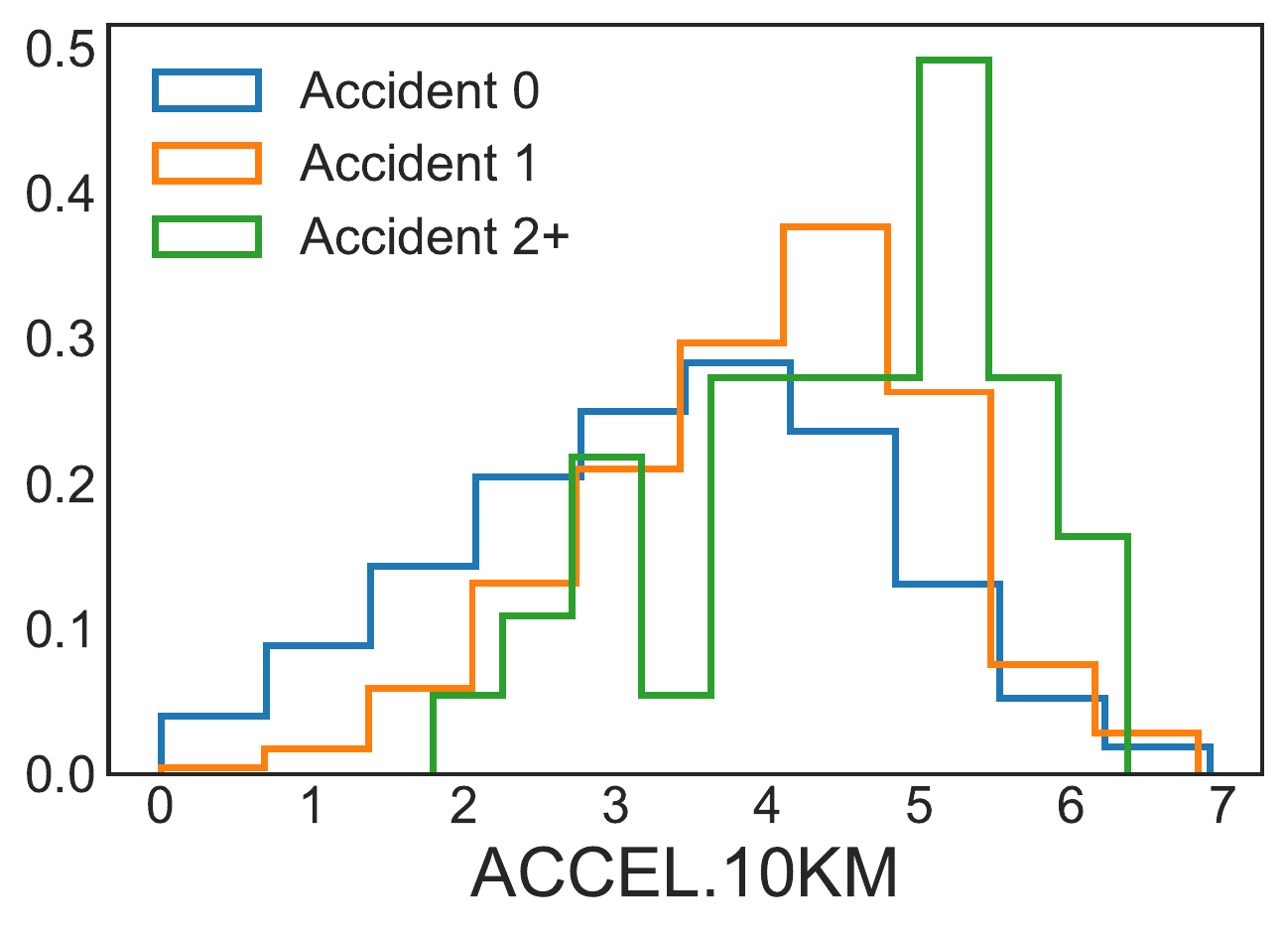} \hfill
  \includegraphics[width=.32\linewidth]{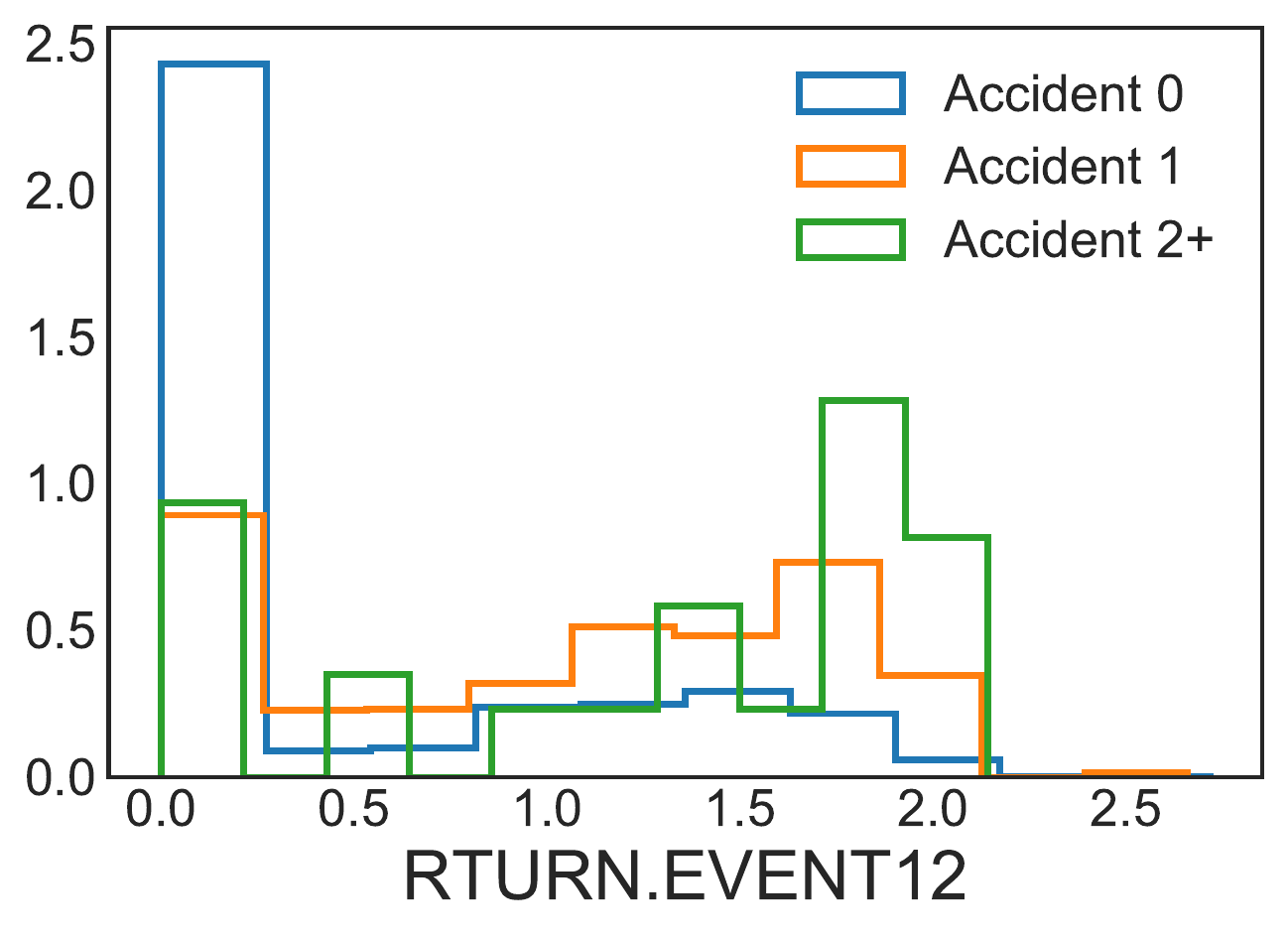} \\
\caption{Histograms of 3 top ranked predictors that are positively related to accident frequency} \label{fig7:pos}
\end{figure}

Finally, considering now all the observations in the training data, we visualize the distribution of the observations according to their accident frequency for the 3 top ranked predictors that are positively related to accident frequency. This is depicted in Figure \ref{fig7:pos}. In each of the sub-figures, we distinguish the accident frequency according to three colors: ACC\_FREQ of 0, 1, and $2^+$ are colored blue, orange, and green respectively. We can somewhat visualize the positive relationship of each of these 3 variables by looking at the right tail of the distribution. For each variable, we observe that the greater the value, the more observations there are in this end of the distribution. All 3 variables are considered driving maneuvers. This demonstrates that while these 3 variables are not necessarily top feature variables ranked by relative importance, the figure exhibits broad positive relationships. The more driving maneuvers the driver violates, the more likely the occurrence of an accident.

\section{Concluding remarks} \label{sec:conclude}

In just the previous 4 to 5 years, we have seen some research work related to vehicle telematics in insurance and actuarial science. This article expands on some of this literature focusing on challenges not previously addressed. In particular, we describe our work that relates to understanding how the addition of telematics information helps improve accuracy for predicting the frequency of accidents. We treated the prediction of accident frequency as a multi-class classification problem with highly imbalanced classes. As with our empirical dataset with telematics, we find that for motor insurance claims portfolio, we typically observe a large number of drivers with zero accidents, a few number with exactly one claim, and far lesser number of drivers with two or more claims. We reviewed some existing multi-class classification models that address the presence of minority class. We find that a combination of resampling procedures and boosting have performed quite well as machine learning tools. We recommend a combination of boosting and cost-sensitive learning algorithms, for which we call \texttt{SAMME.C2}, to address the imbalances within a multi-class classification problem. The injection of the cost-sensitive factors has the effect of placing a heavy penalty to misclassified minority classes and simultaneously heavy reward to correct classifications of minority classes. Results of our model fitting to our telematics dataset are very promising. Conclusively, we demonstrated the superior performance of \texttt{SAMME.C2} with other known boosting algorithms that are also combined with resampling.

When the \texttt{SAMME.C2} algorithm is applied to our telematics dataset, we are able to draw conclusive evidence of how telematics information affects accident frequencies. First, we find that telematics are dominantly better predictors of accident frequencies than just typical classical variables (e.g. DRIVER.AGE, GENDER) used for risk classification. This suggests that telematics variables provide for a better understanding of driver behavior. Second, we are able to group the telematics according to the percentage of time drivers are on the road (e.g. EXPOSURE, PCT.TRIP.TUE, PCT.TRIP.THU) and according to driving maneuvers (e.g. BRAKE.15KM, ACCEL.10KM). We find that broadly speaking, the cluster of variables that describe percentage of time drivers are on the road are relatively more important feature variables than the cluster of variables related to driving maneuvers. Related to this are two observations about driving maneuvers. They are relatively less important because the presence of telematics device in the vehicle is in some indirect sense encouragement of good driving behavior. Drivers can have more control over these driving maneuvers, whereas the time needed to be driving on the road may be more of a necessity (e.g., driving to school or work, driving family members to events, driving for a family vacation). Finally, we find that there is evidence of positive relationships of some of these driving maneuvers to accident frequencies. The more frequent a driver is identified to violate these driving maneuvers, the more likely of the occurrence of accidents. In the future, we wish to investigate the effects of telematics variables on the claim severity or damages associated with an accident.

\bigskip

\subsection*{Appendix A. Detailed steps of the \texttt{SAMME} and \texttt{Ada.C2} algorithms} \label{appa-alg}

\begin{algorithm}[H]
\KwIn{Training dataset $\ \bm{x}_i \in X$, $y_i \in Y = \{1,2,\ldots,K\}$, $T$}
\KwOut{Final classifier $\ H(\bm{x}_i)$}
Set initial distribution of dataset equally distributed:  $D_1(i) = \frac{1}{N}, \quad i =1,2,\ldots,N$ \;
\For{$t=1, \ldots, T$}{
Train weak classifier using the distribution $D_t$\;
Get weak classifier $h_t: X \rightarrow k \in \{1,2,\ldots,K\}$ \;
Compute $\epsilon_t = \dfrac{\sum_{i=1}^{N} D_t(i) I(y_i \ne h_t(\bm{x}_i))}{\sum_{i=1}^N D_t(i)}$ \;
Choose $\alpha_t = \frac{1}{2} \log\!\Big(\dfrac{1-\epsilon_t}{\epsilon_t}\Big) + \log(K-1)$ \;
Update $D_{t+1}(i) = \dfrac{D_t(i) \exp(-\alpha_t I(y_i = h_t(\bm{x}_i)))}{\sum_{j=1}^{N} D_t(j) \exp(-\alpha_t I(y_j = h_t(\bm{x}_j)))}$ \;
}
Return the final classifier: $\ H(\bm{x}_i) = \underset{k}{\mathrm{argmax}} {\  \sum_{t=1}^{T} \alpha_t I(h_t(\bm{x}_i) = k)}$ \;
\caption{\texttt{SAMME}: multi-class AdaBoost}\label{alg:samme}
\end{algorithm}

\begin{algorithm}[htbp]
\KwIn{Training dataset $\ \bm{x}_i \in X$, $y_i \in Y = \{0,1\}$, $\text{Cost}_i$, $T$}
\KwOut{Final classifier $\ H(\bm{x}_i)$}
Set initial distribution of dataset equally distributed:  $D_1(i) = \frac{1}{N}, \quad i =1,2,\ldots,N$ \;
\For{$t=1, \ldots, T$}{
Train weak classifier using the distribution $D_t$\;
Get weak classifier $h_t: X \rightarrow k \in \{0,1\}$ \;
Choose $\alpha_t = \frac{1}{2} \log\! \left(\dfrac{\sum_{i=1}^N I(y_i =h_t(\bm{x}_i)) \, \text{Cost}_i \, D_t(i)}{\sum_{i=1}^N I(y_i \ne h_t(\bm{x}_i)) \, \text{Cost}_i \, D_t(i)} \right)$ \;
Update $D_{t+1}(i) = \dfrac{\text{Cost}_i \, D_t(i) \exp(-\alpha_t I(y_i = h_t(\bm{x}_i)))}{\sum_{j=1}^{N} \text{Cost}_j \, D_t(j) \exp(-\alpha_t I(y_j = h_t(\bm{x}_j)))}$ \;
}
Return the final classifier: $\ H(\bm{x}_i) = \underset{k}{\mathrm{argmax}} {\  \sum_{t=1}^{T} \alpha_t I(h_t(\bm{x}_i) = k)}$ \;
\caption{\texttt{Ada.C2}: cost-sensitive binary AdaBoost}\label{alg:adac2}
\end{algorithm}

\subsection*{Appendix B. Traditional and telematics variables in the dataset} \label{appb-variables}

\begin{table}[H]
\centering
\caption{Variable names and descriptions.} \label{tab:VD}
\resizebox{!}{5.9cm}{
\begin{tabular}{lll}
\\
\toprule
Type & Variable  & Description \\
\midrule
Traditional & DRIVER.AGE  & Age of driver \\
  &  GENDER  & Gender of the driver (M/F) \\
  &  VEHICLE.AGE  & Vehicle age \\
  &  MARITAL  & Marital status \\
  &  VEH.USE  & Use of vehicle: Pleasure, Commute, Farmer, Business  \\
  &  CREDIT.SCORE  & Credit score of driver \\
  &  ZONE  & Zone where driver lives: rural, urban \\
  &  ANN.KMS.DRV.SYSTEM  & Kilometer driven declared by driver \\
  &  YRS.CLAIMS.FREE  & Number of years claims free \\
  &  TERRITORY  & Territory where vehicle is rated \\
\midrule
Telematics & EXPOSURE  & Exposure time in percentage of 24 hours \\
  & DISTANCE.DRIVEN  & Total distance driven \\
  &  PCT.TRIP.xxx  & Percent of driving day xxx of week: MON/TUE/.../SUN \\
  &  PCT.TRIP.xxx  & Percent vehicle driven in xxx hrs: 2HRS/3HRS/4HRS \\
  &  PCT.xxx.DRIV  & Percent vehicle driven in xxx of week: WKDAY/WKEND \\
  &  xx.RUSH.HOUR  & Percent of driving in xx rush hours: AM/PM \\
  &  AVGDAY.USE.WKLY  & Average number of days used per week \\
  &  ACCEL.xxKM  & Number of sudden acceleration 10/13/15.../23 km/h/s per 1000km \\
  &  BRAKE.xxKM  & Number of sudden brakes 10/13/15.../23 km/h/s per 1000km \\
  &  LTURN.EVENTxx  & Number of left turn per 1000km with intensity 08/09/10/11/12 \\
  &  RTURN.EVENTxx  & Number of right turn per 1000km with intensity 08/09/10/11/12 \\
\midrule
Response & ACC\_FREQ  & Frequency of accidents during observation: $0/1/2+$ \\
\bottomrule
\end{tabular}}
\end{table}

\bigskip

\bibliographystyle{apalike}
\bibliography{sammec2.bib}

\begin{thebibliography}{}

\bibitem[Ayuso et~al., 2019]{ayuso2019auto}
Ayuso, M., Guillen, M., and Nielsen, J.~P. (2019).
\newblock Improving automobile insurance ratemaking using telematics:
  incorporating mileage and driver behaviour data.
\newblock {\em Transportation}, 46:735--752.

\bibitem[Ayuso et~al., 2016]{ayuso2016gender}
Ayuso, M., Guillen, M., and P\'{e}rez-Mar\'{i}n, A.~M. (2016).
\newblock Telematics and gender discrimination: some usage-based evidence on
  whether men's risk of accidents differs from women's.
\newblock {\em Risks}, 4:1--10.

\bibitem[Bhowan et~al., 2010]{bhowan2010genetic}
Bhowan, U., Zhang, M., and Johnston, M. (2010).
\newblock Genetic programming for classification with unbalanced data.
\newblock In {\em Proceedings 13th European Conference on Genetic Programming,
  EuroGP 2010}, pages 1--13. Springer-Verlag Berlin.

\bibitem[Boucher et~al., 2017]{boucher2017gam}
Boucher, J.-P., C\^{o}t\'{e}, S., and Guillen, M. (2017).
\newblock Exposure as duration and distance in telematics motor insurance using
  generalized additive models.
\newblock {\em Risks}, 5:1--23.

\bibitem[Chawla et~al., 2002]{chawla2002smote}
Chawla, N.~V., Bowyer, K.~W., Hall, L.~O., and Kegelmeyer, W.~P. (2002).
\newblock {SMOTE}: Synthetic minority over-sampling technique.
\newblock {\em Journal of Artificial Intelligence Research}, 16:321--357.

\bibitem[Chawla et~al., 2003]{chawla2003smboost}
Chawla, N.~V., Lazarevic, A., Hall, L.~O., and Bowyer, K.~W. (2003).
\newblock {SMOTEB}oost: {I}mproving prediction of the minority class in
  boosting.
\newblock In {\em PKDD 2003: Proceedings of the Seventh European Conference on
  Principles and Practice of Knowledge Discovery}, pages 107--119.
  Springer-Verlag: Berlin-Heidelberg.

\bibitem[Constantinescu et~al., 2018]{constantinescu2018impact}
Constantinescu, C.~C., Stancu, I., and Panait, I. (2018).
\newblock Impact study of telematics auto insurance.
\newblock {\em Review of Financial Studies}, 3(4):17--35.

\bibitem[Ferreira and Figueiredo, 2012]{ferreira2012review}
Ferreira, A.~J. and Figueiredo, M.~A. (2012).
\newblock Boosting algorithms: {A} review of methods, theory, and applications.
\newblock In Zhang, C. and Ma, Y., editors, {\em Ensemble Machine Learning:
  Methods and Applications}, chapter~2, pages 35--85. Springer Science.

\bibitem[Fowlkes and Mallows, 1983]{fowlkes1983gmean}
Fowlkes, E.~B. and Mallows, C. (1983).
\newblock A method for comparing two hierarchical clusterings.
\newblock {\em Journal of the American Statistical Association},
  78(383):553--569.

\bibitem[Freund and Schapire, 1997]{freund1997decision}
Freund, Y. and Schapire, R.~E. (1997).
\newblock A decision-theoretic generalization of on-line learning and an
  application to boosting.
\newblock {\em Journal of Computer and System Sciences}, 55(1):119--139.

\bibitem[Friedman et~al., 2000]{friedman2000addlog}
Friedman, J., Hastie, T., and Tibshirani, R. (2000).
\newblock Additive logistic regression: {A} statistical view of boosting.
\newblock {\em The Annals of Statistics}, 28(2):337--407.

\bibitem[Galar et~al., 2012]{galar2012review}
Galar, M., Fern\'{a}ndez, A., Barrenechea, E., Bustince, H., and Herrer, F.
  (2012).
\newblock A review on emsembles for the class imbalance problem: bagging-,
  boosting-, and hybrid-based approaches.
\newblock {\em {IEEE} Transactions on Systems, Man, and Cybernetics -- Part C:
  Applications and Review}, 42(4):463--484.

\bibitem[Guillen et~al., 2019]{guillen2019telem}
Guillen, M., Nielsen, J.~P., Ayuso, M., and P\'{e}rez-Mar\'{i}n, A.~M. (2019).
\newblock The use of telematics devices to improve automobile insurance rates.
\newblock {\em Risk Analysis}, 39(3):662--672.

\bibitem[Guillen et~al., 2020]{guillen2020telem}
Guillen, M., Nielsen, J.~P., P\'{e}rez-Mar\'{i}n, A.~M., and Elpidorou, V.
  (2020).
\newblock Can automobile insurance telematics predict the risk of near-miss
  events?
\newblock {\em North American Actuarial Journal}, 24(1):141--152.

\bibitem[Hastie et~al., 2009]{hastie2009}
Hastie, T., Tibshirani, R., and Friedman, J. (2009).
\newblock {\em The Elements of Statistical Learning: Data Mining, Inference,
  and Prediction}.
\newblock Springer: New York.

\bibitem[Holland, 1975]{holland1975}
Holland, J.~H. (1975).
\newblock {\em Adaptation in Natural and Artifical Systems}.
\newblock Univesity of Michigan Press: Ann Arbor.

\bibitem[Karapiperis et~al., 2015]{cipr2015}
Karapiperis, D., Birnbaum, B., Bradenburg, A., Catagna, S., Greenberg, A.,
  Harbage, R., and Obersteadt, A. (2015).
\newblock Usage-based insurance and vehicle telematics: Insurance market and
  regulatory implications.
\newblock Technical report, National Association of Insurance Commissioners and
  The Center for Insurance Policy and Research.

\bibitem[M\"{u}hlenbein, 1997]{muhlenbein1997GA}
M\"{u}hlenbein, H. (1997).
\newblock Genetic algorithms.
\newblock In Aarts, E.~H. and Lenstra, J.~K., editors, {\em Local Search in
  Combinatorial Optimization}, pages 137--172. Princeton University Press.

\bibitem[Orphanoudakis et~al., 1998]{wong1998}
Orphanoudakis, S.~C., Chronaki, C.~E., Tsiknakis, M., and Kostomanolakis, S.~G.
  (1998).
\newblock Telematics in healthcare.
\newblock In Wong, S.~T., editor, {\em Medical Image Databses}, chapter~10,
  pages 251--281. Springer New York.

\bibitem[Pazzani et~al., 1994]{pazzani1994reduce}
Pazzani, M., Merz, C., Murphy, P., Ali, K., Hume, T., and Brunk, C. (1994).
\newblock Reducing misclassification costs.
\newblock In {\em ICML 1994: Proceedings of the Eleventh International
  Conference on Machine Learning}, pages 217--225. Morgan Kaufman Publishers
  Inc.: San Francisco, CA.

\bibitem[Pednault et~al., 2000]{pednault2000ins}
Pednault, E.~P., Rosen, B.~K., and Apte, C. (2000).
\newblock Handling imbalanced data sets in insurance risk modeling.
\newblock Technical report, Association for the Advancement of Artificial
  Intelligence ({AAAI}).

\bibitem[Pedregosa et~al., 2011]{pedregosa2011scikit}
Pedregosa, F., Varoquaux, G., Gramfort, A., Michel, V., Thirion, B., Grisel,
  O., Blondel, M., Prettenhofer, P., Weiss, R., Dubourg, V., Vanderplas, J.,
  Passos, A., Cournapeau, D., Brucher, M., Perrot, M., and Duchesnay, E.
  (2011).
\newblock Scikit-learn: machine learning in {P}ython.
\newblock {\em Journal of Machine Learning Research}, 12:2825--2830.

\bibitem[P\'{e}rez-Mar\'{i}n et~al., 2019]{perez2019quantile}
P\'{e}rez-Mar\'{i}n, A.~M., Guillen, M., Alca\~{n}iz, M., and Berm\'{u}dez, L.
  (2019).
\newblock Quantile regression with telematics information to assess the risk of
  driving above the posted speed limit.
\newblock {\em Risks}, 7:1--11.

\bibitem[Pesantez-Narvaez et~al., 2019]{pesantez2019xgboost}
Pesantez-Narvaez, J., Guillen, M., and Alca\~{n}iz, M. (2019).
\newblock Predicting motor insurance claims using telematics data -- {XGB}oost
  versus logistic regression.
\newblock {\em Risks}, 7:1--16.

\bibitem[Schapire and Singer, 1999]{schapire1999improv}
Schapire, R.~E. and Singer, Y. (1999).
\newblock Using boosting algorithms using confidence-rated predictions.
\newblock {\em Machine Learning}, 37:297--336.

\bibitem[Seiffert et~al., 2010]{seiffert2010rusboost}
Seiffert, C., Khoshgoftaar, T.~M., {Van Hulse}, J., and Napolitano, A. (2010).
\newblock {RUSB}oost: {A} hybrid approach to alleviating class imbalance.
\newblock {\em IEEE Transactions on Systems, Man, and Cybernetics - Part A:
  Systems and Humans}, 40(1):185--197.

\bibitem[Sun et~al., 2007]{sun2007cost}
Sun, Y., Kamel, M.~S., Wong, A.~K., and Wang, Y. (2007).
\newblock Cost-sensitive boosting for classification of imbalanced data.
\newblock {\em Pattern Recognition}, 40(12):3358--3378.

\bibitem[Verbelen et~al., 2018]{verbelen2018telem}
Verbelen, R., Antonio, K., and Claeskens, G. (2018).
\newblock Unravelling the predictive power of telematics data in car insurance
  pricing.
\newblock {\em Journal of the Royal Statistical Society: Series C (Applied
  Statistics)}, 67(5):1275--1304.

\bibitem[Yang and Wu, 2006]{yang200610probs}
Yang, Q. and Wu, X. (2006).
\newblock {10} challenging problems in data mining research.
\newblock {\em International Journal of Information Technology \& Decision
  Making}, 5(4):597--604.

\bibitem[Zhu et~al., 2009]{zhu2009mclass}
Zhu, J., Zou, H., Rossett, S., and Hastie, T. (2009).
\newblock Multi-class {A}da{B}oost.
\newblock {\em Statistics and Its Interface}, 2:349--360.

\end{thebibliography}

\end{document}